\documentclass[a4paper,12pt,showpacs,showkeys,pre,amsmath,amssymb,floatfix]{revtex4}
\usepackage{graphicx}
\usepackage{xspace}

\newcommand{\csx}{C\ensuremath{_{60}}\xspace}
\newcommand{\ctf}{C\ensuremath{_{240}}\xspace}

\begin{document}

\title{Phase transitions in fullerenes:\\Fragmentation and reassembly of the carbon cage}

\author{Adilah Hussien}
\email{hussein@fias.uni-frankfurt.de}
\affiliation{Frankfurt Institute of Advanced Studies, Johann Wolfgang Goethe---Universit\"at Frankfurt, Ruth-Moufang Stra\ss e 1, D-60438 Frankfurt am Main, Germany}

\author{Alexander Yakubovich}
\email{yakubovich@fias.uni-frankfurt.de}
\affiliation{Frankfurt Institute of Advanced Studies, Johann Wolfgang Goethe---Universit\"at Frankfurt, Ruth-Moufang Stra\ss e 1, D-60438 Frankfurt am Main, Germany}

\author{Andrey Solov'yov}
\altaffiliation[On leave from  ]
{Ioffe Physico-Technical Institute, Polytechnicheskaya 26, St. Petersburg 194021, Russia}
\affiliation{Frankfurt Institute of Advanced Studies, Johann Wolfgang Goethe---Universit\"at Frankfurt, Ruth-Moufang Stra\ss e 1, D-60438 Frankfurt am Main, Germany}

\author{Walter Greiner}
\affiliation{Frankfurt Institute of Advanced Studies, Johann Wolfgang Goethe---Universit\"at Frankfurt, Ruth-Moufang Stra\ss e 1, D-60438 Frankfurt am Main, Germany}

\date{\today}

\begin{abstract}
Phase transition in fullerenes C$_{60}$ and C$_{240}$ are investigated by means of constant-temperature molecular dynamics simulations. In the phase transition region, the assembly (and fragmentation) of the C$_{60}$ cage from (and to) the gaseous state is demonstrated via the dynamical coexistence of two phases. In this critical region, the fullerene system is seen to continuously oscillate between the carbon cage (the solid phase) and the state of carbon dimers and short chains (the gas phase). These oscillations correspond to consecutive disintegration and formation of the fullerene. Furthermore, the temperature-dependent heat capacity of the fullerene features a prominent peak, signifying the finite system analogue of a first-order phase transition. The simulations were conducted for 500~ns using a topologically-constrained pairwise forcefield which was developed for this work. Results of the simulations were supplemented by a statistical mechanics analysis to account for entropy and pressure corrections, corresponding to experimental conditions. These corrections lead to a phase transition temperature of 3800--4200~K for pressure 10--100~kPa, in good agreement with available experimental values.
\end{abstract}



\maketitle

%%%%%%%%%% INTRODUCTION %%%%%%%%%%%%%%
%------------------------------------%
\section{Introduction} %

Thermal fragmentation and assembly of fullerenes can be viewed as reverse processes which have clear features of phase transition. The former leads to the disintegration of the \emph{solid}-like hollow cage to a \emph{gas}-like state of dimers (C$_2$ units); while the latter recreates the fullerene cage from the hot carbon gas. This work investigates the above idea by means of isothermal molecular dynamics (MD) simulations of fullerenes \csx and \ctf. We report the occurrence of fullerene sublimation above a critical temperature---the phase transition temperature---beginning with the loss of a C$_2$ unit. This rapidly leads to the breaking of the fullerene cage and its eventual decomposition into the gaseous phase. In the region of the phase transition temperature, the system oscillates between two distinct phases: the \emph{solid}-like cage and the \emph{gas}-like state of the carbon dimers. Such oscillation corresponds to the consecutive back-and-forth fragmentation and re-assembly of the fullerene which can be seen clearly in the bimodal distribution of the total energy over time (at a single temperature). This coexistence behavior, as well as the prominent peak of the temperature-dependent heat capacity are signatures of first-order phase transition in finite systems \cite{sugano1998}.

Whether the fragmentation is induced by pyrolysis \cite{sommer1996}, laser-irradiation \cite{obrien1988} or collisions
with charged/neutral particles \cite{lutz2002, schmidt2001, ehlich1996}, fullerenes are known to
disintegrate via sequential loss of C$_2$ units, through asymmetric fission or
multifragmentation \cite{rohmund1996,rentenier2003}. Fullerene stability and its fragmentation mechanism
have been investigated extensively by a variety of computational methods, notably tight-binding molecular
dynamics (TBMD). Initial TBMD simulations of the \csx fragmentation were performed by Wang \emph{et al.}
in which the \csx was found to be stable against spontaneous disintegration for temperatures up to 5000~K \cite{wang1992}.
This was followed by similar studies by Zhang \emph{et al.} for fullerenes ranging from C$_{20}$ to C$_{90}$ where,
for small fullerenes (n$\le$58) it was discovered that the fragmentation temperature increased linearly.
However, for larger fullerenes (n=60 and n$\ge$70), the temperature stabilised around 5500~K \cite{zhang1993a,zhang1993b}.
Both studies employed the TB-parametrisation of Xu \emph{et al.} \cite{Xu1992}, as are the work
conducted by L\'aszlo \cite{laszlo1997} and Openov and Podlivaev \cite{openov2006} for the canonical
and microcanonical ensembles respectively. Kim \emph{et. al} reported structural changes
in the \csx and C$_{70}$ in the range of 3000 and 4000~K---with the onset of bond breaking around 5000~K~\cite{kim1993}.
Kim and Tom\`anek conducted fragmentation simulations of the C$_{20}$, \csx and \ctf and
found several different phases of the fullerene melting process, including a liquid-like pretzel
phase \cite{tomanek1994}. While Xu and Scuseria conducted photofragmentation simulations of the \csx and
observed sequential loss of C$_2$ units with cage fragmentation occurring for T$>$5600~K. Horv\`ath and Beu have
also conducted radiation-induced fragmentation of fullerenes and reported multifragmentation to be the main
disintegration channel at high excitation energies. However, between excitation energies of 100--120~eV, they reported
the occurrence of a phase transition where they had defined phase transition to be a steep drop in the average
fragment size with temperature (in their work, the average fragment size dropped from 60 to 5 between 100--120~eV)
\cite{horvath2008}. Semiempirical bond-order methods are also used in studying fullerenes.
The many-body Tersoff potential \cite{tersoff1988} has been applied to a large number of carbon systems,
and has been used by Marcos, \emph{et al.} to investigate the thermal stability of fullerenes \cite{marcos1999}.
The Reactive Bond Order Potential (REBO, or Brenner's potential) \cite{brenner2002}---itself built upon the Tersoff
potential---is also typically employed and has been used to study
the formation of carbon-cage structures \cite{maruyama1998} and it collision-induced
fragmentation \cite{chancey2003}.

Fullerenes can be produced by laser-vaporisation of graphite \cite{kroto1985}, or \emph{en masse} by thermal methods such as electric-arc discharge \cite{kratschmer1990} and the combustion of hydrocarbons in a hot oxygen-rich environment \cite{howard1991, homann1998, mckinnon1992}. They have also been found in candle soot \cite{howard1991}, chinese ink sticks \cite{osawa1997}, areas of asteroid impact \cite{becker1994} and as byproducts of carbon nanotube synthesis \cite{sadana2005}. Despite the technological leaps in the mass production of fullerenes, the detailed atomistic mechanism of fullerene formation has remained elusive and is hampered by the fact that fullerenes are able to self-assemble out of chaotic environments. Nonetheless, a variety of proposed mechanisms exist. Prominent amongst them are: \emph{the Pentagon road} where the fullerene is created by the addition of C$_2$ units or small carbon particles to the dangling bonds of open graphitic cups \cite{smalley1992}; \emph{the Fullerene road} is a similar mechanism to the Pentagon road, however the intermediate structures are small closed-cage fullerenes instead \cite{heath1998}; \emph{Ring Fusion Spiral Zipper} where (instead of C$_2$ units) small carbon rings coalesce to form a cluster, and then ``zip'' up and anneal to assemble into a fullerene \cite{helden1993, hunter1993} and \emph{the Shrinking Hot Giant Fullerene road} where giant fullerenes (with attached dangling chains) are created via heating a gas of dimers at temperatures 2000--3000~K (the size-up process) in a box of side 30~\AA{}. These giant fullerenes are then continuously heated, causing shedding of the side chains and evaporation of C$_2$ units which shrinks the giant fullerenes (the size-down process) to sizes as small as the \csx \cite{irle2006}. Using high-resolution TEM, Huang \emph{et al.} have demonstrated the occurrence of this mechanism where they observed the shrinking of a giant fullerene (C$_{1300}$) inside the cavity of a multi-walled nanotube. The giant fullerene shrinks to the \csx, however, below this size, the cage ruptures open and evaporates \cite{huang2007}. For detailed reviews of the various formation mechanisms, see \cite{smalley1992, singh1995, goroff1996, lozovik1997, morokuma2007}.

In this paper, we address the phase transition properties of \csx and \ctf using molecular dynamics simulations.
We have developed an efficient topologically-constrained pairwise forcefield
that would allow one to simulate these fragmentation/assembly processes on a long time scale (500~ns).
The parameters of the forcefield were chosen in order to reproduce the well-known characteristics of the fullerene molecule.
A simple form of the locally pairwise interaction allowed us to achieve significant computational gains to
observe the coexistence of two phases during the phase transition process.
The results obtained using the developed forcefield are consistent with those obtained with more advanced potentials, see
Table.~\ref{table:fragTemp}. We observe that in a certain temperature range, the fullerene oscillates between two prominently different phases. These states can be considered as solid-like (hollow cage) and gas-like phases (gas of dimers) of the fullerene, and these oscillations correspond to the consecutive fragmentation and reassembly of the fullerene cage at a given temperature. We analyze the dependence of the heat capacity of the system on temperature using two different approaches: based on a) energy fluctuations in the system and b) differentiation of the energy on temperature dependence. Both approaches show that a heat capacity on temperature dependence has a prominent peak which is a signature of a first-order-like phase transition.

The paper is structured as follows. In Section~\ref{methods}, we present the developed forcefield and details of the MD simulations conducted. In Section~\ref{results}, we present and discuss results of the simulations, comparing them to previous works; while in Section~\ref{results:statmech}, we introduce entropic corrections to our results by means of a statistical mechanics analysis and show the correspondence of this theoretical work to experiments. In Section~\ref{conclusions}, we outline the conclusions of this work.

%%%%%%%%%%% METHOD %%%%%%%%%%%%%%
%-------------------------------%
\section{Theoretical and Computational Methods}\label{methods}
\subsection{Topologically-constrained pairwise forcefield}

% FIX: Y
The total energy of an $N$-particle system is given by the Hamiltonian of the form
\begin{equation}\label{eqn:hamiltonian}
H=\sum_{i=1}^{N} \left( \frac{\mathbf{p}_{i}^2}{2m_{i}}+\mathbb{V}(\mathbf{r}_{i}) \right),
\end{equation}
where $\mathbf{r}_{i}$ is the position of atom $i$, $m_i$ its mass and $\mathbf{p}_{i}$ its momentum. The first term represents the kinetic energy, while the second is the potential energy based on the forcefield $V$. This forcefield differentiates between two types of interactions in the system: long-range \emph{van der Waals} (vdW) and short-range \emph{covalent} (cov) bonding. It is expressed as (for all $i\ne j$),
\begin{equation}
\mathbb{V}(\mathbf{r}_{i})= \sum_{j \neq k,l,m}^{N} V_{vdW}\,(r_{ij}) + \sum_{j=k,l,m} V_{cov}\,(r_{ij}),
\end{equation}
where $r_{ij}$ is the interatomic distance between atoms $i$ and $j$ defined
as $r_{ij}=|\mathbf{r}_{i}-\mathbf{r}_{j}|$. The sum in the covalent term is over the three nearest-neighbors of
atom $i$: atoms $k$, $l$ and $m$; while the sum in the van der Waals term is over
the other ($N-4$) non-neighbors of atom $i$.  Hence, each atom is allocated 3 covalent nearest-neighbors
(determined by their positions relative to atom $i$) and $N-4$ vdW non-neighbors. In this way, the atom
is \emph{topologically-constrained} to interact with a certain type of bonding with certain atoms in the system.
%The allocation of each atom and its nearest-neighbors was done by mapping the structure of the \csx (optimized using density functional calculations with RB3LYP functional and the 6-31G(d) basis set).

In this work, for computational efficiency we have chosen the Lennard-Jones type of potential
to treat the interactions between the neighboring atoms although, in principle, one could choose different kind
of pairwise potentials, such as the Morse potential, to model these interactions.
\begin{equation}
V_{LJ}(r_{ij})=\epsilon \left[ \left( \frac{\sigma}{r_{ij}}\right)^{12} - 2\left(\frac{\sigma}{r_{ij}}\right)^{6} \right],
\end{equation}
where $\sigma$ is the equilibrium distance corresponding to the minimum energy $\epsilon$. For the vdW interaction
$V_{vdW}$, the parameters $\epsilon_{vdW}$ and $\sigma_{vdW}$ are based on the nonbonded carbon-carbon interaction
in \cite{mackerell1997}. However, we have  adjusted the value of $\sigma_{vdW}$ to 3.0~\AA{} from 4.0~\AA{} so
that we could reproduce correctly the bond length of \csx.

As for the \emph{covalent} interaction, there are two types of bonds in the \csx: the single and double carbon bonds. Hence, we have two different equilibrium parameters: $\sigma_{s}$ and $\sigma_{d}$ whose corresponding minimum energy parameters are $\epsilon_{s}$ and $\epsilon_{d}$. The values of $\sigma_{s}$ and $\sigma_{d}$ reflect the equilibrium bond lengths of the single and double bonds in the \csx \cite{c60_bondlengths}, while $\epsilon_{s}$ and $\epsilon_{d}$ are based on the carbon-carbon bond dissociation energies in ethane and ethylene respectively \cite{bond_dissoc_energy}. Hence for the \csx,
\begin{equation}\label{cov_c60}
\sum_{j=k,l,m} V_{cov}\,(r_{ij})=V_{s}\,(r_{ik})+V_{s}\,(r_{il})+V_{d}\,(r_{im}),
\end{equation}
since for each atom $i$, there are three neighbors---two of which form the single bond (also known as the 5-6 bond), and the other the double bond (the 6-6 bond). The parameters of the forcefield are summarised in Table~\ref{tab:LJparams}.

\begin{table}[h]
\caption{The fullerene forcefield parameters} \label{tab:LJparams}
\begin{ruledtabular}
\begin{tabular}{c c c c}
Type & $\sigma$ (\AA{}) & $\epsilon$ (eV) & $\epsilon$ (kcal/mol)\\
\hline
Single & 1.45 & 3.81 & 88.00 \\ [1ex]
Double & 1.38 & 6.27 & 144.5 \\ [1ex]
vdW & 3.00 & 0.00052 & 0.0121 \\ [1ex]
\end{tabular}
\end{ruledtabular}
\end{table}

In the case of \ctf, the bonding order is more complicated with bond types intermediate between the single and the double bond. To fit the parameters for this (and all other types of fullerenes aside from the \csx), we first calculate all the bond lengths of the DFT-optimized \ctf and then set these lengths as the equilibrium distance $\sigma_{ij}$ for atom $i$ and its neighbor $j$. To calculate the minimum energy parameters, we constructed the following scaling
\begin{equation}
\epsilon_{ij}=\frac{\sigma_{s}-\sigma_{ij}}{\sigma_{s}-\sigma_{d}}(\epsilon_{d}-\epsilon_{s}) + \epsilon_{s},
\end{equation}
which returns the minimum energy parameters for the \csx if $\sigma_{ij}=\sigma_{s}$ or $\sigma_{ij}=\sigma_{d}$ corresponding to (\ref{cov_c60}).

\subsection{Molecular dynamics simulations}
With this forcefield, we have performed constant-temperature molecular dynamics (MD) simulations for \csx and \ctf. The time evolution of the system was obtained by integrating numerically the Newtonian equations of motion using the leapfrog technique \cite{leapfrog}. The time step of the simulation is $\triangle t$=1 fs, with each run (at each temperature) being 500~ns long ($5\times10^{8}$ steps in total). We have assumed the first 500~ps for equilibration purposes, thus have only considered data from the remaining trajectory in our analysis. The interaction cutoff was set to 15~\AA{} and the simulation was performed using periodic boundary conditions with a unit cell of length 20~\AA{} per side. In contrast to previous simulations that use simple velocity-scaling \cite{zhang1993a} or the Nos\'e-Hoover thermostat \cite{tomanek1994}, temperature control was achieved in this work by using Langevin Dynamics which acts as a stochastic heat bath \cite{kampen1981,bussi2007}. The advantages of Langevin Dynamics are: a) it is known to generate a canonical ensemble (unlike velocity-scaling) and b) it simulates the heat transfer of the particles in the heat bath through physical ``collisions and friction'' processes (unlike the Nos\'e-Hoover thermostat) \cite{chelikowsky1996}. It is described, for each particle $i$, as
\begin{equation}
m_{i}\frac{d^2r_{i}}{dt^2}=F_{i}(r)+R_{i}(t)-m_{i}\gamma_{i}\frac{dr_{i}}{dt},
\end{equation}
where $\gamma_{i}$ is the friction coefficient set to 100~ps$^{-1}$ and $R_{i}(t)$ is the random force vector that simulates noise in the system. It is defined as a stationary Gaussian process with zero mean: $\langle R_{i}(t)R_{j}(t') \rangle=2m_{i}\gamma_{i}kT\delta(t')\delta_{ij}$, where $k$ is the Boltzmann constant, T the temperature and $\delta$ is the Dirac delta function.

The simulation was conducted for temperatures between 3000--8500~K. In the phase transition range, we have performed
runs in temperature steps of 5~K; while outside this range, in steps of 25~K. Simultaneously,
we have also varied the $\epsilon_{s}$ parameters of the \csx ($\epsilon_{s}$=2.38, 3.25, 3.81, 4.12 and 4.99~eV)
to investigate its effect on the phase transition temperature.
In addition, we have employed a static neighbors list throughout the entire simulation, \emph{i.e.}
we have specified the three nearest-neighbors for each carbon atom at the beginning of the
simulation and it is only with these neighbors that each atom is able to `form' and `break' bonds.
We discuss the influence of this restriction on the dynamics of the system in Sec.~\ref{results:statmech}.

%%%%%%%%%%% RESULTS & ANALYSIS %%%%%%%%%%%%%%
%-------------------------------------------%
\section{Results and Analysis}\label{results}
\subsection{Caloric curves}

\begin{figure}[!ht]
\includegraphics[scale=0.65]{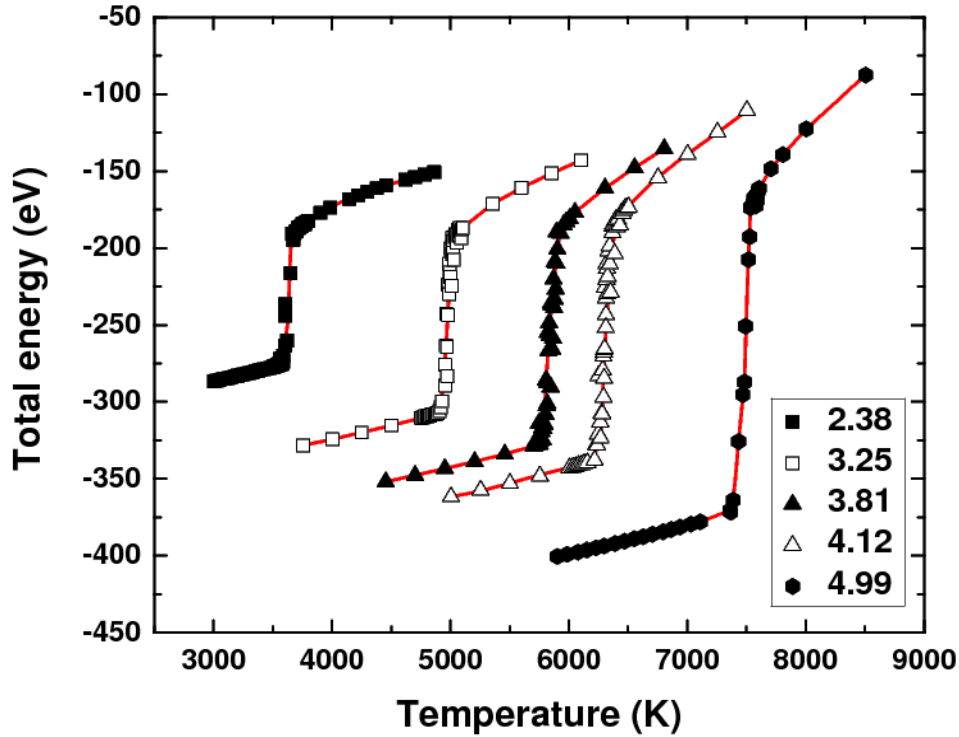}
\caption{Total energy $E$ of \csx as a function of temperature for $\epsilon_{s}$=2.38, 3.25, 3.81, 4.12 and 4.99~eV. Each curve shows a distinct jump in energy corresponding to a phase transition in the system. The scattered plot is the time-average total energy and the solid thick line is the cubic B-spline interpolation.} \label{fig:energycurves}
\end{figure}

We present, in Fig.~\ref{fig:energycurves}, the temperature dependence of the total energy,as expressed in Eq. (\ref{eqn:hamiltonian}), for five different $\epsilon_{s}$ parameters ($\epsilon_{s}$=2.38, 3.25, 3.81, 4.12 and 4.99~eV). Each of these curves can be divided into three distinct parts corresponding to three different phases of the fullerene: the region before the abrupt jump in energy (the solid-like phase), the region after it (the gas-like phase) and the region of the jump itself (the phase transition). As can be observed, an increase in the $\epsilon_{s}$ energy parameter is correlated with an increase in both the temperature at which the phase transition occurs and the height of the phase transition barrier. This behavior is due to more energy being required to pull apart the bonds in the fullerene. We have also conducted simulations where the $\epsilon_{d}$ energy parameters are varied for a given $\epsilon_{s}$ parameter which is held constant at 3.81~eV. However, this does not affect significantly the thermal behavior of the fullerene as only 60 out of the 180 bonds in the \csx are double bonds.

The energy curves corresponding to $\epsilon_{s}$ parameters 3.25~eV and 3.81~eV (equivalent to the C-C bond energy in ethylene) show that an energy value of 100--140~eV is required for the jump from one phase to another. This result is in very good agreement with that of Horv\'ath and Beu who had conducted tight-binding MD simulations and found a multfragmentation phase transition occurring for energies between 100--120~eV \cite{horvath2008}. It is also in align with previous results from other constant temperature \cite{tomanek1994,laszlo1997} and constant energy \cite{marcos1999,scuseria1994} MD simulations which report a result in the range of 90--96~eV.

\begin{figure}[!ht]
\includegraphics[scale=0.6]{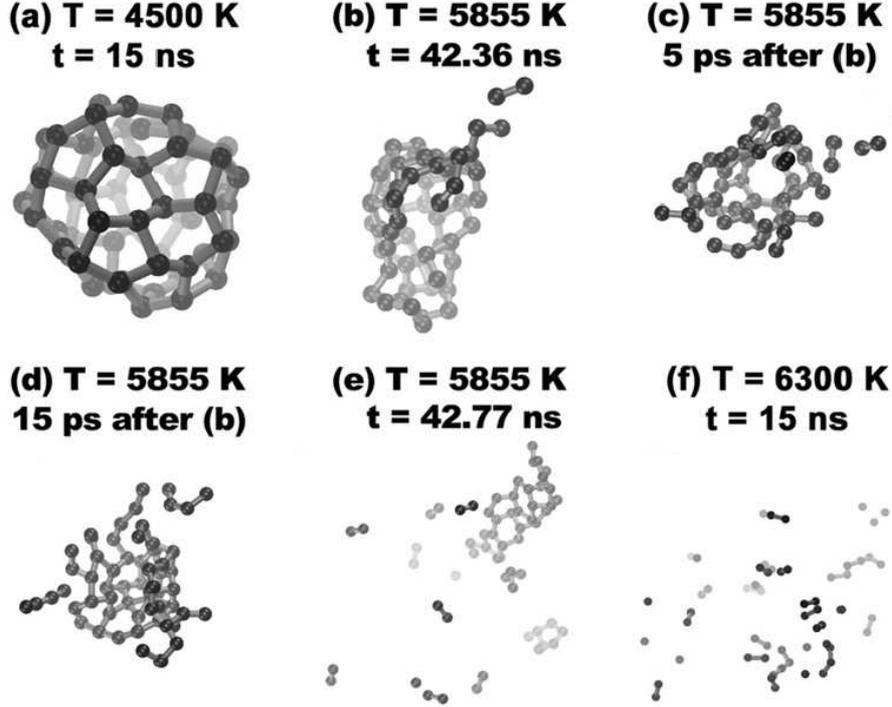}
\caption{Snapshots of the fragmentation simulation of \csx. (a) Deformed \csx cage at T=4500~K; (b)
Broken bonds and evaporation of a C$_{2}$ from the cage at T=5855~K; (c) The broken cage starts unravelling
and continues evaporating more dimers T=5855~K, 5~ps after the initial C$_{2}$ evaporation;
(d) Cage breaks into a web-structure and short chains at T=5855~K;
(e) Web-structure decomposes to smaller chains and dimers at T=5855~K; (f) Only dimers and short
chains at T=6300~K.} \label{fig:snapc60}
\end{figure}

To illustrate the structural changes that occur in the fullerene, leading to a phase transition, we present in Fig.~\ref{fig:snapc60} snapshots of the simulation process for $\epsilon_{s}$=3.81~eV. The fullerene is found in its \emph{solid}-like phase at T $\lesssim$ 5500~K, as demonstrated in Fig.~\ref{fig:energycurves} by the linearly increasing region before the jump. In Fig.~\ref{fig:snapc60}(a), the fullerene can be seen to be intact although structurally deformed at T=4450~K. At T=5855~K, the carbon bonds are observed to be broken and fragmentation occurs with the onset of a C$_{2}$ evaporation, Fig.~\ref{fig:snapc60}(b). This causes the unravelling of the cage structure, as more dimers, short chains and rings are evaporated, Fig.~\ref{fig:snapc60}(c-e). At T $\gtrsim$ 5855~K, the fullerene is found to be in its \emph{gas}-like phase of dimers and short chains. As demonstrated in our simulations, the fragmentation/evaporation of the fullerene occurs before the development of a liquid-like \cite{kim1993}, pretzel or linked-chain \cite{tomanek1994} phases. Instead, the fullerene is seen to decompose from a stable solid-like cage structure to a stable gas-like phase of dimers and short chains, with no stable intermediate structure of long carbon chains that would signify a liquid-like phase. This result is also observed by Marcos, et al. who attribute the liquid-like phases to insufficient simulation time \cite{marcos1999}.

\subsection{The coexistence regime}
\begin{figure*}[ht]
\begin{center}
\includegraphics[scale=0.65]{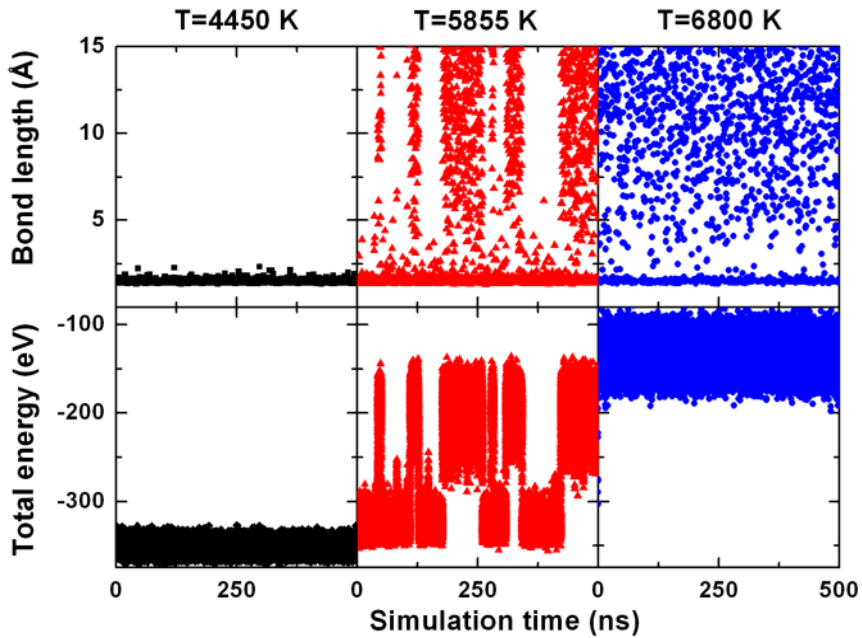}
\caption{\emph{Top}: Bond length as a function of simulation time for a single bond for trajectories with $\epsilon_{s}$=3.81~eV. At 4450~K (left), the bonds average about 1.5~\AA{}. At 5855~K (middle), abrupt jumps in the bond length can be seen. These correspond to phase transitions from the solid-like cage phase to the gas-like phase of dimers and short chains. At 6800~K (right), the homogeneous distribution of bond lengths is characteristic of the gas-like phase. \emph{Bottom}: Total Energy, $E$, as a function of simulation time over the entire trajectory, including the equilibration process, at T=4450, 5855 and 6800~K.} \label{fig:osc}
\end{center}
\end{figure*}

In Fig.~\ref{fig:osc}, we present the bond length of a characteristic single bond in the fullerene as a function of simulation time (for $\epsilon_{s}$=3.81~eV). The figure spans the whole trajectory (500~ns) including the equilibration process which we have taken to be the first 500~ps of the simulation. At 4450~K, the system is located in the region \emph{before} the phase transition, Fig.~\ref{fig:energycurves}, and correspondingly, the bond lengths do not fluctuate more than 2.0~\AA{}, indicating that although the fullerene might be vibrating and deformed in shape, the cage-like structure is still present. At 6800~K, the system is located in the region \emph{after} the phase transition, and the homogeneous distribution of the bond lengths is characteristic of a gas-like phase. At 5855~K however, one can observe abrupt columns of homogeneous distributions occurring for intervals in the simulation. These regions correspond to periods in the trajectory where the fullerene has jumped into a \emph{gas}-like phase. This on-and-off distribution is characteristic of a two-phase coexistence regime that has been observed for finite-size clusters (see \cite{sugano1998} and references therein).

\begin{figure*}[hp]
\begin{center}
\includegraphics[scale=0.55]{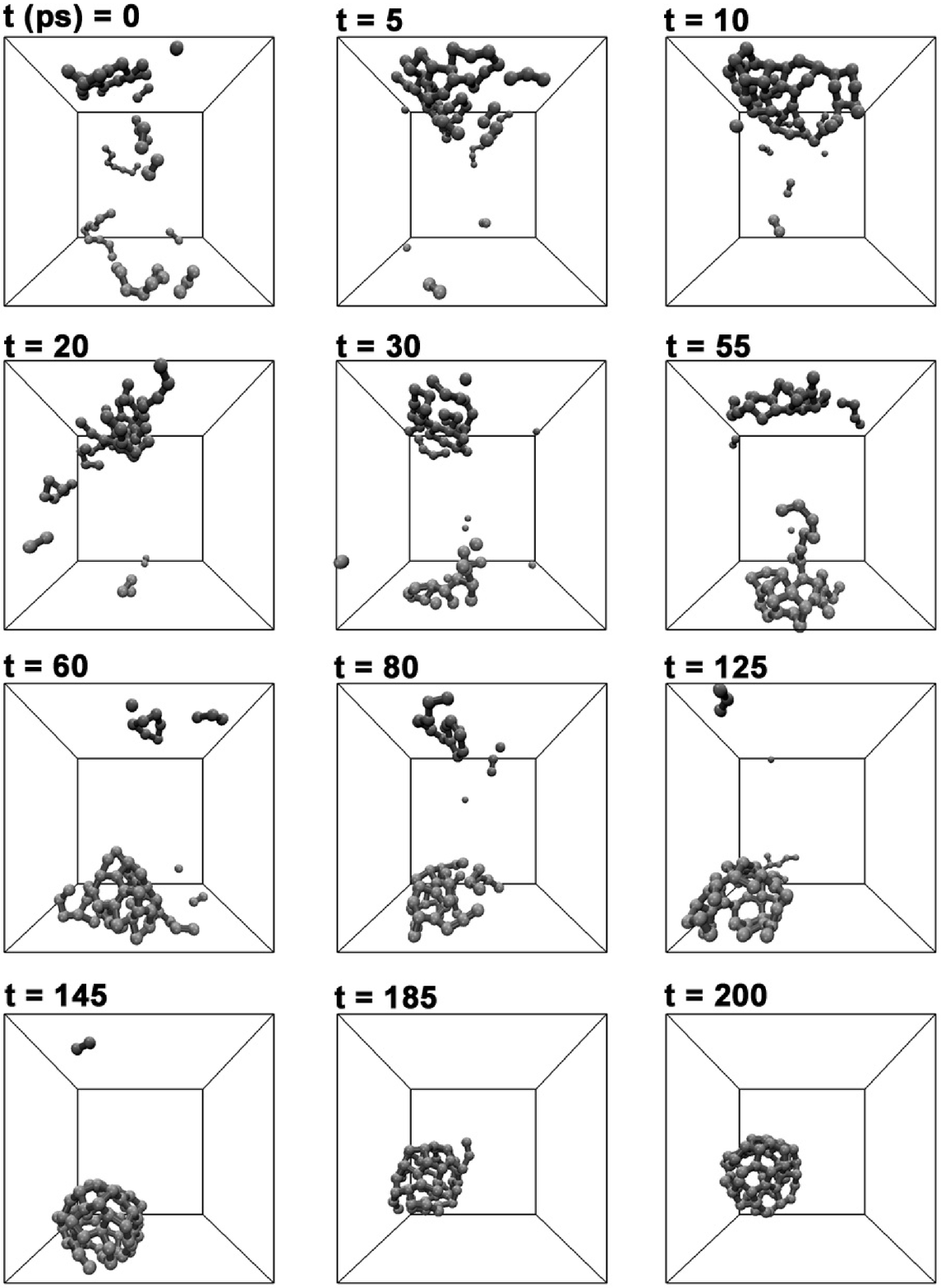}
\caption{Snapshots of the re-assembly process (gas-like phase to solid-like phase) from the coexistence behavior of \csx in the phase transition region at T=5855~K. All times are in ps and denote the interval from the first snapshot.} \label{fig:coex}
\end{center}
\end{figure*}

In Fig.~\ref{fig:osc}, we plot the corresponding time-dependent total energy for T=5855~K. Here, one can observe the abrupt energy jumps between the solid-like phase ($\sim$-325~eV) and the gas-like phase ($\sim$-200~eV) throughout the entire trajectory. This behavior has not been observed in previous studies on fullerene fragmentation and thermal stability which we attribute to the short time scales of these simulations, as they are usually on the order of 10 ps to 1 ns, whereas our simulations are on the order of hundreds of $ns$. In Fig.~\ref{fig:coex}, we present snapshots of the transition from a gas-like phase of dimers and short chains, to the solid-like phase of a carbon cage. In other words, this is a re-assembly/formation process of the fullerene cage where dimers and small chains link up to form a discernible web-like structure that grows into the hollow cage. The cage however, is quite deformed as one would expect at such a high temperature, nonetheless it is an obvious fullerene cage.

\subsection{Heat capacity and phase transition temperature}

\begin{figure}[!ht]
\begin{center}
\includegraphics[scale=0.65]{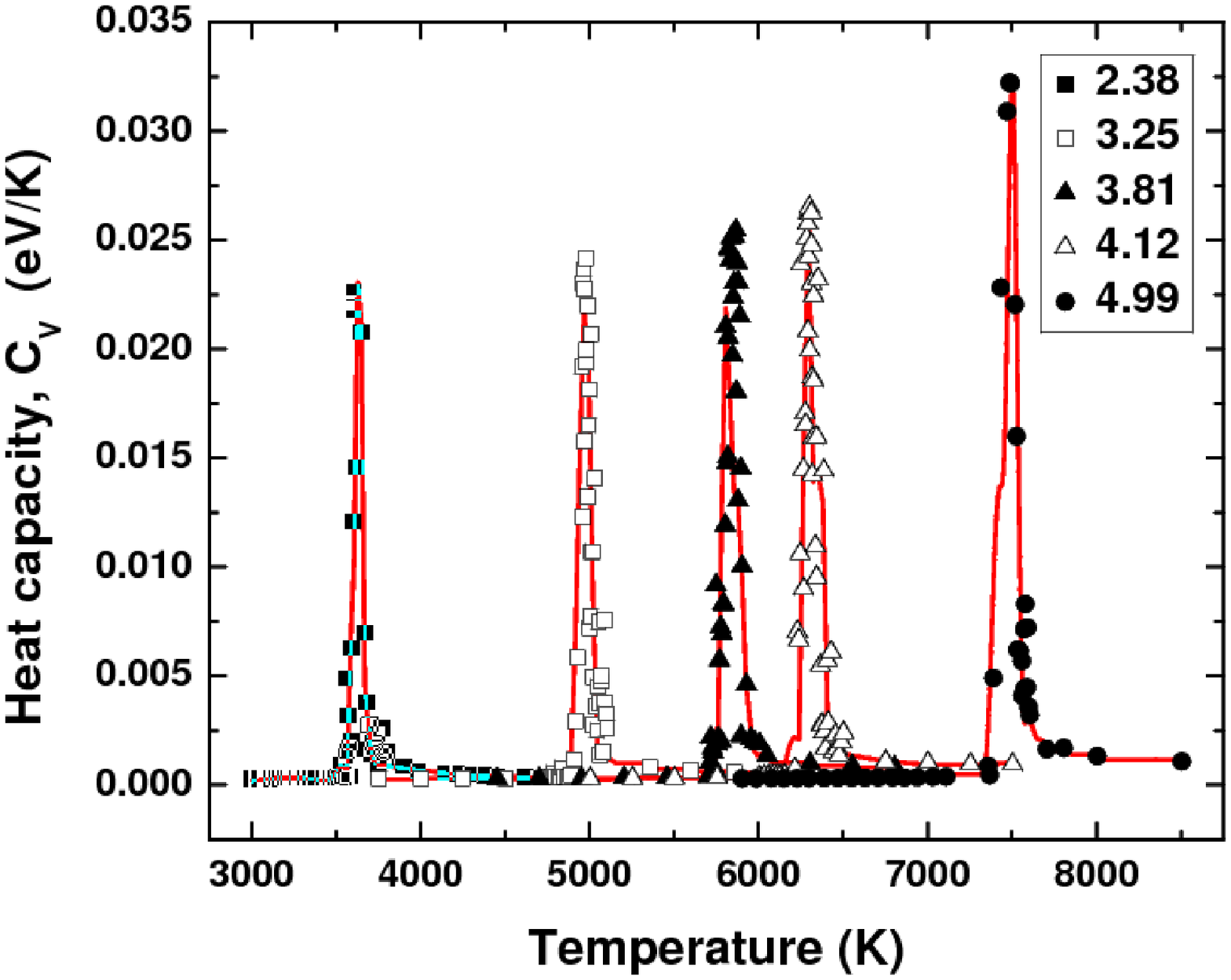}
\caption{\csx heat capacity plots obtained from $c_{v}=dE/dT$ for different $\epsilon_{s}$ parameters. The temperature at the maximum of each heat capacity curve is denoted the phase transition temperature, T$_{PT}$: 3500~K (2.38~eV), 4950~K (3.25~eV), 5855~K (3.81~eV), 6450~K (4.12~eV), 7450~K (4.99~eV). The scatter plots are calculated according to Eq. (\ref{flucheatcap}), while the thick solid lines were differentiated from the cubic B-spline interpolation from Fig.~\ref{fig:energycurves}.} \label{fig:heatcap}
\end{center}
\end{figure}

\begin{table}[h]
\caption{Entropy change for the different $\epsilon_{s}$ parameters.} \label{table:heatcapentropy}
\begin{ruledtabular}
\begin{tabular}{ c c c}
$\epsilon_{s} (eV)$ & T$_{PT} (K)$ & $\Delta S$ \\
\hline
2.38 & 3500 & 0.025\\ [1ex]
3.25 & 4950 & 0.024\\ [1ex]
3.81 & 5855 & 0.024\\ [1ex]
4.12 & 6450 & 0.025\\ [1ex]
4.99 & 7450 & 0.027\\ [1ex]
% [5ex]
\end{tabular}
\end{ruledtabular}
\end{table}

In Fig.~\ref{fig:heatcap}, we have plotted the heat capacity of the \csx obtained from $c_{v}=dE/dT$, where $E$ is the total energy at temperature $T$. The solid lines in Fig.~\ref{fig:heatcap} were generated by initially interpolating the associated energy curve using cubic B-splines (as can be seen in the thick solid lines in Fig.~\ref{fig:energycurves}) and then by differentiating these interpolated lines. The scattered data points were calculated using the energy fluctuations in the system \cite{leach2001}:
\begin{eqnarray}\label{flucheatcap}
c_{v}&=&\frac{\langle E^2 \rangle-\langle E \rangle^2}{kT^2}\\
&=&\frac{\langle\,(E-\langle E \rangle\,)\,^2\, \rangle}{kT^2},
\end{eqnarray}
where $\langle E \rangle$ is the time-average total energy at temperature $T$ and $k$ is the Boltzmann constant. The maximum of each heat capacity curve denotes the temperature at which the phase transition (PT) is defined to occur. We note that in previous papers, the fragmentation temperature had also been defined differently from this work. They were taken to indicate the onset of bond-breaking \cite{zhang1993a, zhang1993b, wang1992} or the dissociation of a C$_{2}$ unit from the fullerene cage \cite{openov2006,laszlo1997, scuseria1994}. These differences are, in part, due to different simulation times and the potentials/methods used. In Table~\ref{table:fragTemp}, we have arranged the values of the fragmentation (or PT) temperature to compare our results with previous works.

\begin{table}[h]
\caption{Fragmentation temperature of \csx} \label{table:fragTemp}
\begin{ruledtabular}
\begin{tabular}{ c c c}
Model & T$_{f}$ & Ref. \\
\hline
Zhang, et al. & 5500 & \cite{zhang1993a} \\ [1ex]
Wang, et al. & 5500 & \cite{wang1992} \\ [1ex]
Xu \& Scuseria & 5600 & \cite{scuseria1994} \\ [1ex]
Kim \& Toman\`ek & 4000 & \cite{tomanek1994} \\ [1ex]
Kim, Young \& Lee & 4800 & \cite{kim1993} \\ [1ex]
L\`aszl\`o & 6800 & \cite{laszlo1997} \\ [1ex]
This model & 5855 & - \\ [1ex]
\end{tabular}
\end{ruledtabular}
\end{table}

We note that the maximum values of our heat capacity plots are of an order of magnitude higher than available fragmentation studies with published heat capacity data \cite{tomanek1994,laszlo1997,kim1993}. This is clearly due to the sharp jump found to occur in the total energy plots due to the long simulations times of this work. The almost constant heat capacity distribution across the $\epsilon_{s}$ parameters show that the entropy change, $\Delta S$ from the \emph{solid}-like phase to the \emph{gas}-like phase is almost constant as
\begin{equation}\label{entropyeqn}
\Delta S=\frac{\Delta E}{T_{0}},
\end{equation}
where $\Delta S$ is the change in entropy, $\Delta E$ is the change in energy from one phase to the other (see Fig.~\ref{fig:energycurves}) and T$_{0}$ is the phase transition temperature. In Table~\ref{table:heatcapentropy}, we have arranged the entropy change, $\Delta S$, corresponding to the different single carbon bond energy parameters, $\epsilon_{s}$ and the phase transition temperature, T$_{0}$.

\subsection{Statistical mechanics and correspondence to experiment}
\label{results:statmech}

The focus of this work is the phase transition dynamics of fullerene-like systems. We have constructed
a fullerene model based on a topologically-constrained forcefield which allows for gain in the simulation time.
To further speed up the simulation process, we have only employed a static neighbors list throughout the entire simulation,
\emph{i.e.} we have specified the three nearest-neighbors for each carbon atom at the beginning
of the simulation and it is only with these neighbors that each atom is able to `form' and `break' bonds.
In other words, we have labeled the bonding between atoms throughout the entire simulation.

While such bond-numbering is reasonable when the system is in the cage structure, it is less 
so when the system moves towards the phase transition region and in its gaseous state. 
This specific bond-numbering prevents the system from forming new bonds. 
Thus, each carbon atom will only seek out its three allocated neighbors that were specified at the beginning of 
the simulation. In other words, our system has been constrained to only one particular combination of 
bonding that would lead to the fullerene cage. However, for a system consisting of 
30 \emph{indistinguishable} dimers, there should be $30!\,2^{30}$ possible combinations 
available for the system, all of which would lead to the same cage structure. This correction has to 
be taken into account in order to specify the correct phase transition temperature. Additionally, 
pressure corrections must also be accounted. In typical electric arc discharge experiments, 
the gas pressure in such setups vary between 13 to 70~kPa \cite{huczko1997,saidane2004,akita2000,markovic2003}; 
however, the pressure in our simulation box is much higher ($\sim$0.4~GPa) due to 
its small volume.

\squeezetable
\begin{table}[hb]
\caption{Partition of the energy at T=5855~K} \label{table:equipartition}
\begin{ruledtabular}
\begin{tabular}{c c c}
Type & Cage phase energy (eV) & Gas phase energy (eV) \\
\hline
Single bonds & -200.18 & -68.82 \\ [1ex]
Double bonds & -177.51 & -173.67 \\ [1ex]
van der Waals & +7.045 & + 2.399 \\ [1ex]
\hline
Total & -370.65 & -240.09 \\ [1ex]
\end{tabular}
\end{ruledtabular}
\end{table}

In order to introduce the necessary corrections,
let us first analyze the partition of the energy in the system at T=5855~K
(the phase transition temperature) when it is in the cage and gas states (see Table~\ref{table:equipartition}).
The difference in the total potential energy when the cage is at T=5855~K and when the system is optimized
is equal to 45.8~eV. This value is equal to $\frac{3}{2}$NkT (where N=60 is the total number of atoms,
k is the Boltzmann constant and T=5855~K), being in correspondence with the statistical mechanics
description of our system.
Table~\ref{table:equipartition} shows also that
the remaining single bonds energy in the gas phase is equal to -68.82~eV.
This energy contribution is due to the continuous formation and fragmentation of unstable short chains, thus the system in our simulation
does not remain entirely in a phase of 30 dimers. However, in the following statistical mechanics description,
we assume that the system can reach the gaseous state of 30 entirely unbound C$_2$ dimers
at temperatures close to the phase transition temperature.

One can now calculate the entropy change of the system undergoing a phase transition with both
bond numbering and pressure corrections taken into account.
The entropy change of the system experiencing a phase transition can be defined as $\Delta S=S_{gas}-S_{cage}$,
where $S_{gas}$ is the entropy of the gaseous state and $S_{cage}$ is the entropy of the cage structure of the fullerene.
Since there are $2^{30}\,30!$ ways to assemble a fullerene from 30 dimers,
the term k$\ln(2^{30}\,30!)$ should be added to the entropy of the cage structure.
The entropy of a dimer being in the gaseous phase is proportional to the logarithm of the volume accessible to it.
Therefore the entropy correction to the gaseous state of 30 statistically independent carbon dimers
in some arbitrary volume (for instance, the volume $V_{exp}$ corresponding to experimental conditions) is equal
to k$\ln\left[(\frac{V_{exp}}{V_{sim}})^{30}\right]=30$k$\ln\left[\frac{V_{exp}}{V_{sim}}\right]$ or, in terms of
concentrations $30$k$\ln\left[\frac{n_{sim}}{n_{exp}}\right]$. The similar contribution accounting for the
entropy of the cage structure within the given volume should be added to $S_{cage}$. Thus, the total entropy change of the system
with the corrections has the following form:

\begin{equation}\label{indistinguishable}
\Delta S=\Delta S_{0}-k\ln(2^{30}\,30!)+30k\ln\left[\frac{n_{sim}}{n_{exp}}\right]-k\ln\left[\frac{n_{sim}}{n_{exp}}\right],
\end{equation}

\noindent where $k$ is Boltzmann constant and $\Delta S_{0}=\frac{\Delta E}{T_{0}}$ is the entropy change
of the system experiencing a phase transition from a fullerene cage to the gas state in a simulation box of
volume 8000~\AA$^3$. Here the energy variation
$\Delta E=\Delta E_0+P\Delta V$ represents the energy difference between the cage structure and
the system in the gas phase. We have assumed that the short chains of carbon dimers fragment almost at the 
temperature of the phase transition, therefore $\Delta E_0$=200.18~eV is the energy difference
between the cage state and the state where no single bonds are present. 
This value was obtained from molecular dynamics simulations performed at the temperature 5855~K. However, for any arbitrary temperature $\Delta E_0$ can be calculated with approximately 5\% accuracy as follows: 
$\Delta E_0(T)=E_{sb}-\frac{5}{2}$kT. The first term in the right hand side of the expression 
$E_{sb}=60\epsilon_s=228.6$~eV is the total energy of single bonds for the equilibrated structure 
(see Table~\ref{tab:LJparams}) and the second term is the potential part of the vibrational energy of 
the dimers within the cage.
The term $P \Delta V=(N_{gas}-N_{cage})k T_0=29$kT$_0$ is the contribution to  $\Delta E$ which arises due to the presence
of pressure in the system. This term accounts for the energy needed to overcome the external pressure in the system during the process of fullerene
fragmentation.
%We assume that the mean free path of the dimers in the gas phase in our simulation is much smaller than the size of the simulation box.}
T$_{0}=$5855~K is the phase transition temperature obtained from the simulations while $n_{sim}$
and $n_{exp}$ are the concentrations of carbon dimers in the simulation and at experimental conditions respectively.
The second term in the right hand side of Eq. (\ref{indistinguishable}) accounts for the permutation correction of the entropy
in the cage state as discussed above.
The third term represents the entropy change
of the gas of 30 carbon dimers with variation of the system volume from $V_{sim}$ to $V_{exp}$,
while the fourth term is a similar variation but for the gas of fullerene molecules instead.
We have assumed that, in the gas phase, the dimers are statistically independent.
Hence, Eq. (\ref{indistinguishable}) can be rewritten as:

\begin{equation}\label{indistinguishable2}
\Delta S=\frac{\Delta E}{T_{0}}-k\ln(2^{30}\,30!)+29k\ln n_{sim}-29k\ln\left[\frac{P}{kT}\right],
\end{equation}

\noindent where $P=n_{exp}kT$ is the pressure in the system.
Here $n_{sim}^{-1}=(V_{box}-V_{excl})/30$, where $V_{box}$=8000~\AA$^3$ is the volume of the simulation box
and $V_{excl}$ is the excluded volume due to the van der Waals repulsion between the atoms.
The excluded volume is estimated through the distance at which the repulsion energy between dimers is equal to kT. 
This distance $r(T)$ is calculated as a root of the following equation:

\begin{equation}\label{eqn:repulsion}
\epsilon \left[ \left( \frac{\sigma}{r}\right)^{12} - 2\left(\frac{\sigma}{r}\right)^{6} \right]=kT,
\end{equation}

\noindent where $\epsilon$ and $\sigma$ are the parameters of the forcefield given in Table~\ref{tab:LJparams}.
Knowing the dependence of r on temperature and the fact that each pair of atom 
are not allowed to be closer than the distance r, one can calculate the excluded volume:  
$V_{excl}(T)= 30\cdot\frac{4 \pi}{3}r^3$, which at T=5855~K, is equal to  $V_{excl}$=822~\AA$^3$.

\begin{figure}[!ht]
\begin{center}
\includegraphics[scale=0.9]{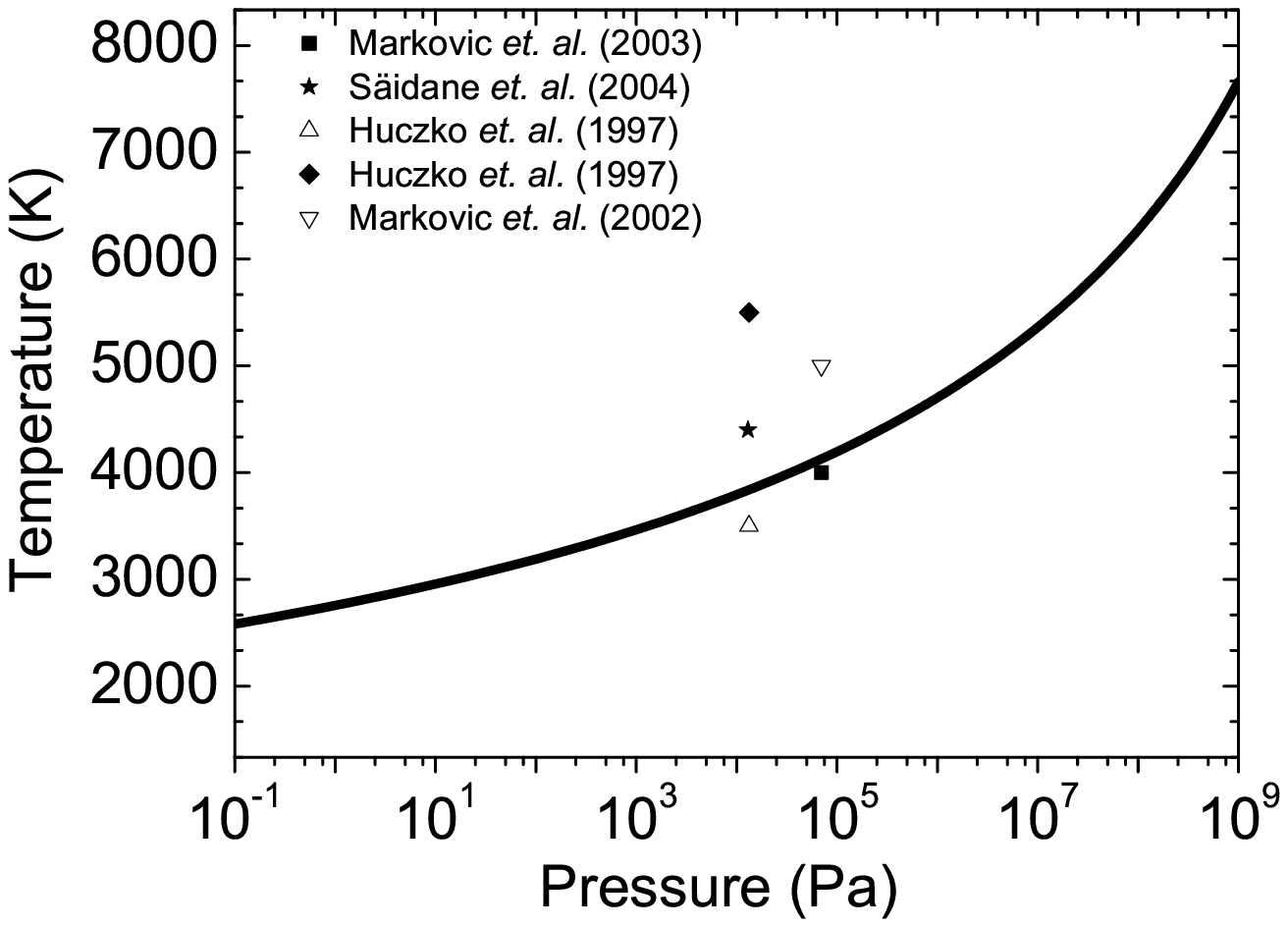}
\caption{Dependance of \csx phase transition temperature on pressure in the gas phase is plotted by solid line.
Data points in the figure refer
to different carbon-arc experimental conditions: $\blacksquare$ Markovic \emph{et al.} \cite{markovic2003},
$\bigstar$ S\"aidane \emph{et al.} \cite{saidane2004}, $\bigtriangleup$ and $\blacklozenge$ Huczko \emph{et al.}
\cite{huczko1997}, $\bigtriangledown$ Markovic \emph{et al.} \cite{markovic2002}.}

\label{fig:pressure}
\end{center}
\end{figure}

Knowing $\Delta S$ as a function of pressure, one can then evaluate the phase transition temperature as
a root of the following expression:
\begin{equation}\label{eqn:ph_temperature}
T= \frac{ E_{sb}-\frac{5}{2}kT\cdot30+P\Delta V(T)}{\Delta S(T)},
\end{equation}

\noindent  where $E_{sb}=60\epsilon_s=228.6$~eV is the total energy of the single bonds 
for the equilibrated structure (see Table~\ref{tab:LJparams}), while the second term
in the nominator is the potential part of vibrational energy of the dimers in the cage.
These two terms describe the energy difference between the cage and the gaseous state of the fullerene
at the phase transition temperature.
The third term $P \Delta V(T)=29$kT accounts for
the energy necessary for increasing the volume of the system during the fragmentation process. 
The dependence of the phase transition temperature on pressure is presented in Fig.~\ref{fig:pressure}.

From Fig.~\ref{fig:pressure}, it is seen that for the pressure range of $10--100$~kPa 
the phase transition temperature is between 3800--4200~K, being in a good agreement 
with the temperatures reported in the arc discharge experiments \cite{saidane2004,markovic2003}. 
%depending on experimental parameters such as the current intensity, electrode diameter and the gas flow rate.
It is important to note that our statistical mechanics model is developed for 
for an thermodynamically equilibrated  system. However, in electric arc discharge experiments the condition of local thermal equilibrium for 
the arc region can be justified due to high current density ($\sim$320 A/cm$^2$) and high pressures (40--70~kPa) \cite{akita2000}. 
Fig.~\ref{fig:pressure} shows
that with lowering the pressure, the phase transition temperature decreases.
However, at these conditions, the equilibration time  increases exponentially with the decrease 
of the phase transition temperature and at some point it should exceed the
available experimental time limits.

\subsection{The \ctf}
\begin{figure}[!ht]
\begin{center}
\includegraphics[scale=1]{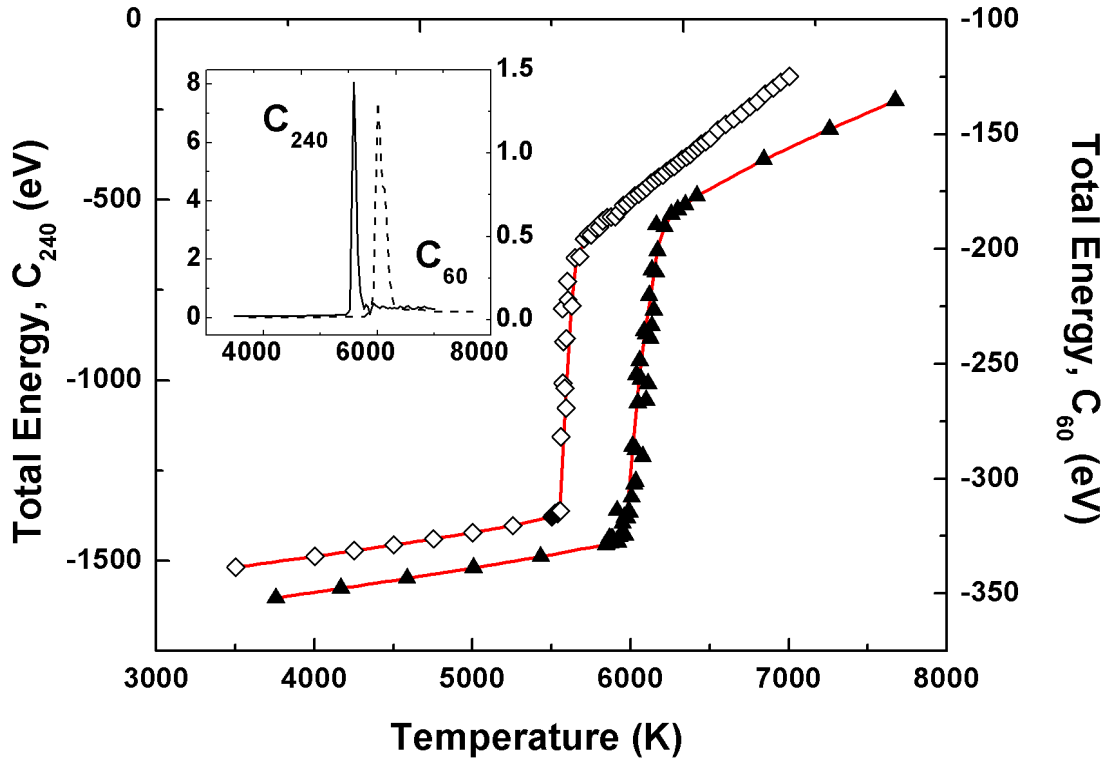}
\caption{Total energy 
of \ctf as a function of temperature compared to that 
for  \csx. \emph{Open} squares correspond to \ctf and \emph{closed} triangles to \csx. 
The left vertical scale corresponds to \ctf and the right to \csx. \emph{Inset}: heat capacity plots 
for \ctf and \csx, the horizontal axis is temperature in K and the vertical axes 
are the heat capacity in~eV/K.} \label{fig:c240energy}
\end{center}
\end{figure}

\begin{figure}[!ht]
\begin{center}
\includegraphics[scale=0.6]{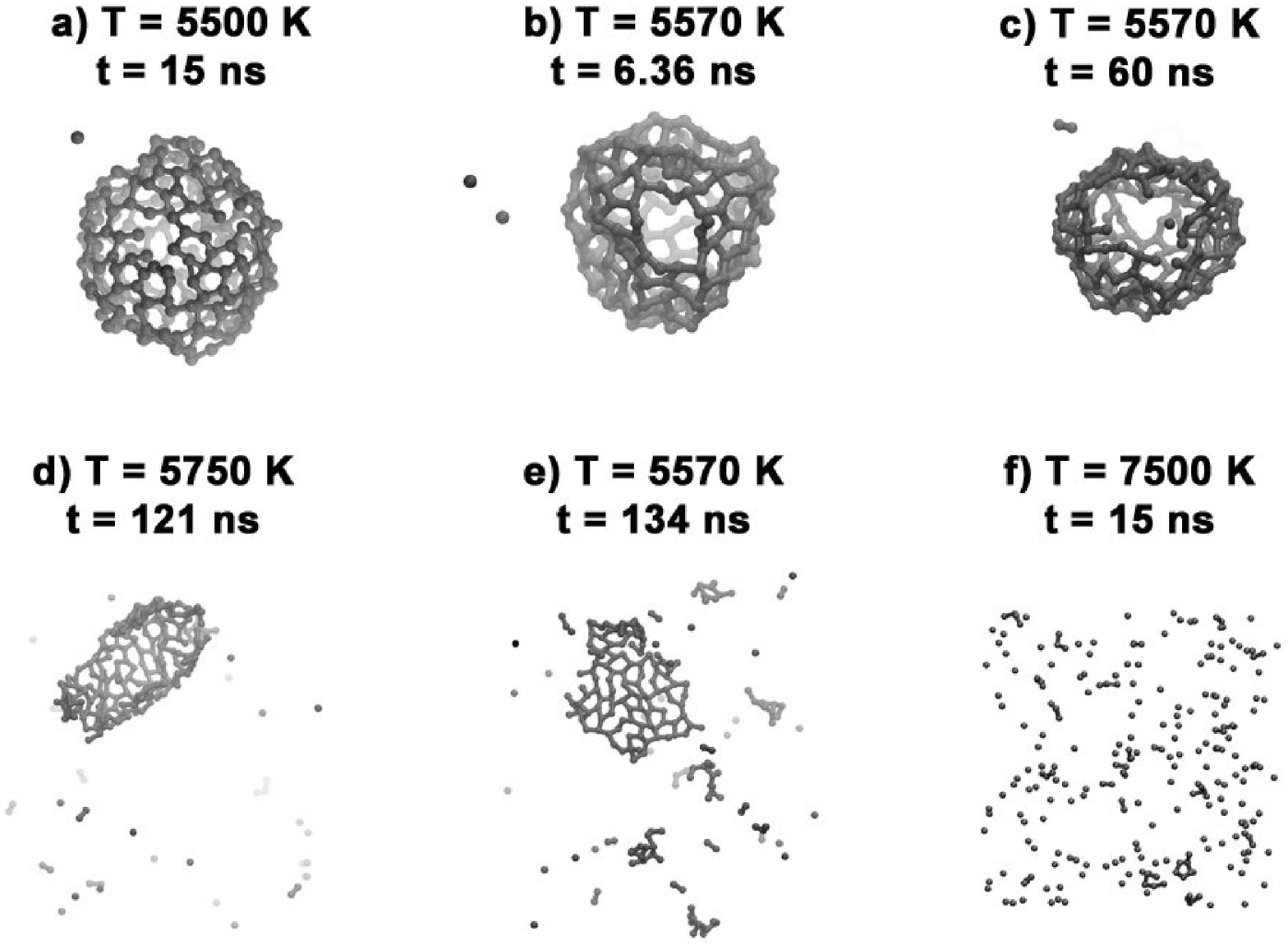}
\caption{Snapshots of the fragmentation simulation of \ctf: (a) T=5500~K, distorted fullerene cage evaporates of a C unit, (b) T=5570~K, broken bonds and formation of hole in cage corresponding to further evaporation of C units, (c) T=5570~K, evaporation of a dimer, (d-f) fragmentation of the cage follows from the onset of bond-breaking and evaporation until complete dimerization in (f) at T=7500~K. \emph{Note}: All times in the figure correspond to the simulation time at the \emph{particular} temperature only.} \label{fig:snapc240}
\end{center}
\end{figure}

In Fig.~\ref{fig:c240energy} we have plotted the total energy  of 
\ctf as a function of temperature and compared it with that for \csx. 
As shown, the phase transition of the \ctf occurs at a lower 
temperature T $\sim$ 5500~K, while the energy difference between the solid-like and 
gas-like phases of the \ctf is $\sim$750~eV---five times larger than that required 
for phase transition of the \csx. To illustrate this process, we have included the snapshots 
of the \ctf fragmentation process in Fig.~\ref{fig:snapc240}. Similar to the \csx, the \ctf also undergoes 
deformation of the cage which eventually leads to the evaporation of a dimer as the first step towards 
complete disintegration. However, we observe that the \ctf is more readily to form larger 
fragments, chains and rings and, unlike the \csx, it does not reach the coexistence regime at the 
the phase transition temperature. This regime for \ctf is expected to happen at a longer simulation times.
As before, we have derived the heat capacity 
for the \ctf, see inset to Fig.~\ref{fig:c240energy}, and from its peak location we have determined 
the fragmentation temperature being equal to be 5500~K. Note that in both figures, the vertical scale 
on the right-hand side corresponds to \csx while the left-hand side to \ctf.

We have demonstrated from our simulations, the fragmentation process of the fullerene using our constructed forcefield. 
Our simulation does not show \emph{liquid}-like phases of the fullerene and the phase transition 
is seen to occur between the \emph{solid}-like and \emph{gas}-like phases.
The solid-to-liquid transition are not thermally stable and will fragment into a gas-like 
phase if the simulation is run long enough \cite{zhang1993a,marcos1999}. Most studies in 
this area have been performed with MD simulations that are between 10 ps to 1 ns, although it is 
known experimentally that both laser-induced and thermal fragmentation of fullerenes occur on the 
$\mu$s timescale \cite{kroto1985,obrien1988,kratschmer1990}. This therefore implies that since the 
time scale of any simulation is much shorter than the scale of the experiments, the MD fragmentation temperatures 
will therefore overestimate the experimental ones. However, longer simulation time should allow 
one to observe phases that are stable in the fullerene fragmentation process, and in this field, 
our simulations are two orders of magnitude longer than other MD simulations performed. 
Nonetheless, one must also take into account that in our model, the carbon atoms are unable to 
form bonds with atoms that have not been designated as its neighbors, and this might affect both 
the observation a liquid-like phase and the energetics of the fragmentation process. 
However, our result of a phase transition occurring at T=3800--4200~K and requiring an 
energy of 130--140~eV is in good agreement with previous work in this field \cite{horvath2008,marcos1999,tomanek1994}. 
We have also shown a novel coexistence behavior of the \csx in the phase transition region
and have observed the reassembly 
process of the fullerene cage from a gas of carbon dimers.

%%%%%%%%%% CONCLUSIONS %%%%%%%%%%%%%%
%-----------------------------------%
\section{Conclusions}\label{conclusions}
We have conducted constant-temperature molecular dynamics simulations of \csx and \ctf. Without pressure and entropy corrections, our simulations demonstrate that both \csx and \ctf undergo a phase transition to a \emph{gas}-like state at 5855~K and 5500~K respectively. For the \csx we report a two-phase coexistence behavior where the fullerene continuously oscillates between a \emph{solid}-like hollow cage and a \emph{gas}-like state of carbon dimers. Such oscillation corresponds to continuous fragmentation and reassembly of the fullerene cage (see Figs.~\ref{fig:snapc60} and~\ref{fig:coex}). We demonstrate that the process of fullerene fragmentation and reassembly is a first-order-like phase transition between gaseous and cage states in the finite system. The developed topologically-constrained forcefield allowed for simulations that are 500~ns long, more than two orders of magnitude longer than those conducted in previous works and the results using this forcefield correlate well with those from more sophisticated models (See table~\ref{table:fragTemp}). Using statistical mechanics arguments, we have shown that in the absence of bond-numbering and taking into account pressure corrections to correspond to experimental conditions, the phase transition temperature of \csx would be found in the range of 3800--4200~K (corresponding to pressures between 10--100~kPa). We have also demonstrated a good correspondence of our model to experimental conditions in the arc discharge method (See Fig.~\ref{fig:pressure}).

%%%%%%%%%% ACKNOWLEDGEMENT %%%%%%%%%%
%-----------------------------------%
\begin{acknowledgements}
The authors thank Drs. I.~A.~Solov'yov, E.~S.~Henriques and A.~G.~Lyalin for helpful
advice and guidance. Author A.~H. would like to thank M.~Mathew, B.~Harutyunyan, H.~Fatehi and
S.~Lo for computational help and useful discussions. Our simulations were made possible by the Frankfurt Center for Scientific Computing.
This work was supported by the European Commission within PECU and the Network of Excellence projects. 
\end{acknowledgements}

%%%%%%%%%% BIBLIOGRAPHY %%%%%%%%%%
%--------------------------------%


\begin{thebibliography}{54}
\expandafter\ifx\csname natexlab\endcsname\relax\def\natexlab#1{#1}\fi
\expandafter\ifx\csname bibnamefont\endcsname\relax
  \def\bibnamefont#1{#1}\fi
\expandafter\ifx\csname bibfnamefont\endcsname\relax
  \def\bibfnamefont#1{#1}\fi
\expandafter\ifx\csname citenamefont\endcsname\relax
  \def\citenamefont#1{#1}\fi
\expandafter\ifx\csname url\endcsname\relax
  \def\url#1{\texttt{#1}}\fi
\expandafter\ifx\csname urlprefix\endcsname\relax\def\urlprefix{URL }\fi
\providecommand{\bibinfo}[2]{#2}
\providecommand{\eprint}[2][]{\url{#2}}

\bibitem[{\citenamefont{Sugano and Koizumi}(1998)}]{sugano1998}
\bibinfo{author}{\bibfnamefont{S.}~\bibnamefont{Sugano}} \bibnamefont{and}
  \bibinfo{author}{\bibfnamefont{H.}~\bibnamefont{Koizumi}},
  \emph{\bibinfo{title}{Microcluster Physics}} (\bibinfo{publisher}{Springer},
  \bibinfo{year}{1998}), \bibinfo{edition}{2nd} ed.

\bibitem[{\citenamefont{Sommer et~al.}(1996)\citenamefont{Sommer, Kruse, and
  Roth}}]{sommer1996}
\bibinfo{author}{\bibfnamefont{T.}~\bibnamefont{Sommer}},
  \bibinfo{author}{\bibfnamefont{T.}~\bibnamefont{Kruse}}, \bibnamefont{and}
  \bibinfo{author}{\bibfnamefont{P.}~\bibnamefont{Roth}}, \bibinfo{journal}{J.
  Phys. B: At. Mol. Opt. Phys.} \textbf{\bibinfo{volume}{29}},
  \bibinfo{pages}{4955} (\bibinfo{year}{1996}).

\bibitem[{\citenamefont{O'Brien et~al.}(1988)\citenamefont{O'Brien, Heath,
  Curl, and Smalley}}]{obrien1988}
\bibinfo{author}{\bibfnamefont{S.~C.} \bibnamefont{O'Brien}},
  \bibinfo{author}{\bibfnamefont{J.~R.} \bibnamefont{Heath}},
  \bibinfo{author}{\bibfnamefont{R.~F.} \bibnamefont{Curl}}, \bibnamefont{and}
  \bibinfo{author}{\bibfnamefont{R.~E.} \bibnamefont{Smalley}},
  \bibinfo{journal}{J. Chem. Phys.} \textbf{\bibinfo{volume}{88}},
  \bibinfo{pages}{220} (\bibinfo{year}{1988}).

\bibitem[{\citenamefont{Reink\"oster et~al.}(2002)\citenamefont{Reink\"oster,
  Siegmann, Werner, Huber, and Lutz}}]{lutz2002}
\bibinfo{author}{\bibfnamefont{A.}~\bibnamefont{Reink\"oster}},
  \bibinfo{author}{\bibfnamefont{B.}~\bibnamefont{Siegmann}},
  \bibinfo{author}{\bibfnamefont{U.}~\bibnamefont{Werner}},
  \bibinfo{author}{\bibfnamefont{B.~A.} \bibnamefont{Huber}}, \bibnamefont{and}
  \bibinfo{author}{\bibfnamefont{H.~O.} \bibnamefont{Lutz}},
  \bibinfo{journal}{J. Phys. B: At. Mol. Opt. Phys.}
  \textbf{\bibinfo{volume}{35}}, \bibinfo{pages}{4989} (\bibinfo{year}{2002}).

\bibitem[{\citenamefont{Kunert and Schmidt}(2001)}]{schmidt2001}
\bibinfo{author}{\bibfnamefont{T.}~\bibnamefont{Kunert}} \bibnamefont{and}
  \bibinfo{author}{\bibfnamefont{R.}~\bibnamefont{Schmidt}},
  \bibinfo{journal}{Phys. Rev. Lett.} \textbf{\bibinfo{volume}{86}},
  \bibinfo{pages}{5258} (\bibinfo{year}{2001}).

\bibitem[{\citenamefont{Ehlich et~al.}(1996)\citenamefont{Ehlich, Westerburg,
  and Campbell}}]{ehlich1996}
\bibinfo{author}{\bibfnamefont{R.}~\bibnamefont{Ehlich}},
  \bibinfo{author}{\bibfnamefont{M.}~\bibnamefont{Westerburg}},
  \bibnamefont{and} \bibinfo{author}{\bibfnamefont{E.~E.~B.}
  \bibnamefont{Campbell}}, \bibinfo{journal}{J. Chem. Phys.}
  \textbf{\bibinfo{volume}{104}}, \bibinfo{pages}{1900} (\bibinfo{year}{1996}).

\bibitem[{\citenamefont{Rohmund et~al.}(1996)\citenamefont{Rohmund, Glotov,
  Hansen, and Campbell}}]{rohmund1996}
\bibinfo{author}{\bibfnamefont{F.}~\bibnamefont{Rohmund}},
  \bibinfo{author}{\bibfnamefont{A.~G.} \bibnamefont{Glotov}},
  \bibinfo{author}{\bibfnamefont{K.}~\bibnamefont{Hansen}}, \bibnamefont{and}
  \bibinfo{author}{\bibfnamefont{E.~E.~B.} \bibnamefont{Campbell}},
  \bibinfo{journal}{J. Phys. B: At. Mol. Opt. Phys.}
  \textbf{\bibinfo{volume}{29}}, \bibinfo{pages}{5143} (\bibinfo{year}{1996}).

\bibitem[{\citenamefont{Rentenier et~al.}(2003)\citenamefont{Rentenier,
  Bordenave-Montesquieu, Moretto-Capelle, and
  Bordenave-Montesquieu}}]{rentenier2003}
\bibinfo{author}{\bibfnamefont{A.}~\bibnamefont{Rentenier}},
  \bibinfo{author}{\bibfnamefont{D.}~\bibnamefont{Bordenave-Montesquieu}},
  \bibinfo{author}{\bibfnamefont{P.}~\bibnamefont{Moretto-Capelle}},
  \bibnamefont{and}
  \bibinfo{author}{\bibfnamefont{A.}~\bibnamefont{Bordenave-Montesquieu}},
  \bibinfo{journal}{J. Phys. B: At. Mol. Phys.} \textbf{\bibinfo{volume}{36}},
  \bibinfo{pages}{1585} (\bibinfo{year}{2003}).

\bibitem[{\citenamefont{Wang et~al.}(1992)\citenamefont{Wang, Xu, Chan, and
  Ho}}]{wang1992}
\bibinfo{author}{\bibfnamefont{C.~Z.} \bibnamefont{Wang}},
  \bibinfo{author}{\bibfnamefont{C.~H.} \bibnamefont{Xu}},
  \bibinfo{author}{\bibfnamefont{C.~T.} \bibnamefont{Chan}}, \bibnamefont{and}
  \bibinfo{author}{\bibfnamefont{K.~M.} \bibnamefont{Ho}}, \bibinfo{journal}{J.
  Phys. Chem.} \textbf{\bibinfo{volume}{96}}, \bibinfo{pages}{3563 }
  (\bibinfo{year}{1992}).

\bibitem[{\citenamefont{Zhang et~al.}(1993{\natexlab{a}})\citenamefont{Zhang,
  Wang, Chan, and Ho}}]{zhang1993a}
\bibinfo{author}{\bibfnamefont{B.~L.} \bibnamefont{Zhang}},
  \bibinfo{author}{\bibfnamefont{C.~Z.} \bibnamefont{Wang}},
  \bibinfo{author}{\bibfnamefont{C.~T.} \bibnamefont{Chan}}, \bibnamefont{and}
  \bibinfo{author}{\bibfnamefont{K.~M.} \bibnamefont{Ho}},
  \bibinfo{journal}{Phys. Rev. B} \textbf{\bibinfo{volume}{48}},
  \bibinfo{pages}{11381} (\bibinfo{year}{1993}{\natexlab{a}}).

\bibitem[{\citenamefont{Zhang et~al.}(1993{\natexlab{b}})\citenamefont{Zhang,
  Wang, Ho, and Chan}}]{zhang1993b}
\bibinfo{author}{\bibfnamefont{B.~L.} \bibnamefont{Zhang}},
  \bibinfo{author}{\bibfnamefont{C.~Z.} \bibnamefont{Wang}},
  \bibinfo{author}{\bibfnamefont{K.~M.} \bibnamefont{Ho}}, \bibnamefont{and}
  \bibinfo{author}{\bibfnamefont{C.~T.} \bibnamefont{Chan}},
  \bibinfo{journal}{Z. Phys. D} \textbf{\bibinfo{volume}{26}},
  \bibinfo{pages}{285} (\bibinfo{year}{1993}{\natexlab{b}}).

\bibitem[{\citenamefont{Xu et~al.}(1992)\citenamefont{Xu, Wang, T., and
  Ho}}]{Xu1992}
\bibinfo{author}{\bibfnamefont{C.}~\bibnamefont{Xu}},
  \bibinfo{author}{\bibfnamefont{C.~Z.} \bibnamefont{Wang}},
  \bibinfo{author}{\bibfnamefont{C.~C.} \bibnamefont{T.}}, \bibnamefont{and}
  \bibinfo{author}{\bibfnamefont{K.~M.} \bibnamefont{Ho}}, \bibinfo{journal}{J.
  Phys.: Condens. Matt.} \textbf{\bibinfo{volume}{4}}, \bibinfo{pages}{6047}
  (\bibinfo{year}{1992}).

\bibitem[{\citenamefont{L\'aszl\'o}(1997)}]{laszlo1997}
\bibinfo{author}{\bibfnamefont{I.}~\bibnamefont{L\'aszl\'o}},
  \bibinfo{journal}{Fullerenes, Nanotubes, Carbon Nanostruct.}
  \textbf{\bibinfo{volume}{5:2}}, \bibinfo{pages}{375} (\bibinfo{year}{1997}).

\bibitem[{\citenamefont{Openov and Podliavev}(2006)}]{openov2006}
\bibinfo{author}{\bibfnamefont{L.~A.} \bibnamefont{Openov}} \bibnamefont{and}
  \bibinfo{author}{\bibfnamefont{A.~I.} \bibnamefont{Podliavev}},
  \bibinfo{journal}{JETP Lett.} \textbf{\bibinfo{volume}{84}},
  \bibinfo{pages}{68} (\bibinfo{year}{2006}).

\bibitem[{\citenamefont{Kim et~al.}(1993)\citenamefont{Kim, Lee, and
  Lee}}]{kim1993}
\bibinfo{author}{\bibfnamefont{E.}~\bibnamefont{Kim}},
  \bibinfo{author}{\bibfnamefont{Y.~H.} \bibnamefont{Lee}}, \bibnamefont{and}
  \bibinfo{author}{\bibfnamefont{J.~Y.} \bibnamefont{Lee}},
  \bibinfo{journal}{Phys. Rev. B} \textbf{\bibinfo{volume}{48}},
  \bibinfo{pages}{18230} (\bibinfo{year}{1993}).

\bibitem[{\citenamefont{Kim and Tom\'anek}(1994)}]{tomanek1994}
\bibinfo{author}{\bibfnamefont{S.~G.} \bibnamefont{Kim}} \bibnamefont{and}
  \bibinfo{author}{\bibfnamefont{D.}~\bibnamefont{Tom\'anek}},
  \bibinfo{journal}{Phys. Rev. Lett.} \textbf{\bibinfo{volume}{72}},
  \bibinfo{pages}{2418} (\bibinfo{year}{1994}).

\bibitem[{\citenamefont{Horv\'ath and Beu}(2008)}]{horvath2008}
\bibinfo{author}{\bibfnamefont{L.}~\bibnamefont{Horv\'ath}} \bibnamefont{and}
  \bibinfo{author}{\bibfnamefont{T.~A.} \bibnamefont{Beu}},
  \bibinfo{journal}{Phys. Rev. B} \textbf{\bibinfo{volume}{77}},
  \bibinfo{pages}{075102} (\bibinfo{year}{2008}).

\bibitem[{\citenamefont{Tersoff}(1988)}]{tersoff1988}
\bibinfo{author}{\bibfnamefont{J.}~\bibnamefont{Tersoff}},
  \bibinfo{journal}{Phys. Rev. Lett.} \textbf{\bibinfo{volume}{61}},
  \bibinfo{pages}{2879} (\bibinfo{year}{1988}).

\bibitem[{\citenamefont{Marcos et~al.}(1999)\citenamefont{Marcos, Alonso,
  Rubio, and L\'opez}}]{marcos1999}
\bibinfo{author}{\bibfnamefont{P.~A.} \bibnamefont{Marcos}},
  \bibinfo{author}{\bibfnamefont{J.~A.} \bibnamefont{Alonso}},
  \bibinfo{author}{\bibfnamefont{A.}~\bibnamefont{Rubio}}, \bibnamefont{and}
  \bibinfo{author}{\bibfnamefont{M.~J.} \bibnamefont{L\'opez}},
  \bibinfo{journal}{Eur. Phys. J. D} \textbf{\bibinfo{volume}{6}},
  \bibinfo{pages}{221} (\bibinfo{year}{1999}).

\bibitem[{\citenamefont{Brenner et~al.}(2002)\citenamefont{Brenner, Shenderova,
  Harrison, Stuart, and Sinnott}}]{brenner2002}
\bibinfo{author}{\bibfnamefont{D.~W.} \bibnamefont{Brenner}},
  \bibinfo{author}{\bibfnamefont{O.~A.} \bibnamefont{Shenderova}},
  \bibinfo{author}{\bibfnamefont{J.~A.} \bibnamefont{Harrison}},
  \bibinfo{author}{\bibfnamefont{S.~J.} \bibnamefont{Stuart}},
  \bibnamefont{and} \bibinfo{author}{\bibfnamefont{S.~B.}
  \bibnamefont{Sinnott}}, \bibinfo{journal}{J. Phys.: Condens. Matter}
  \textbf{\bibinfo{volume}{14}}, \bibinfo{pages}{783} (\bibinfo{year}{2002}).

\bibitem[{\citenamefont{Yamaguchi and Maruyama}(1998)}]{maruyama1998}
\bibinfo{author}{\bibfnamefont{Y.}~\bibnamefont{Yamaguchi}} \bibnamefont{and}
  \bibinfo{author}{\bibfnamefont{S.}~\bibnamefont{Maruyama}},
  \bibinfo{journal}{Chem. Phys. Lett.} \textbf{\bibinfo{volume}{286}},
  \bibinfo{pages}{336} (\bibinfo{year}{1998}).

\bibitem[{\citenamefont{Chancey et~al.}(2003)\citenamefont{Chancey, Oddershede,
  Harris, and Sabin}}]{chancey2003}
\bibinfo{author}{\bibfnamefont{R.~T.} \bibnamefont{Chancey}},
  \bibinfo{author}{\bibfnamefont{L.}~\bibnamefont{Oddershede}},
  \bibinfo{author}{\bibfnamefont{F.~E.} \bibnamefont{Harris}},
  \bibnamefont{and} \bibinfo{author}{\bibfnamefont{J.~R.} \bibnamefont{Sabin}},
  \bibinfo{journal}{Phys. Rev. A} \textbf{\bibinfo{volume}{67}},
  \bibinfo{pages}{043203} (\bibinfo{year}{2003}).

\bibitem[{\citenamefont{Kroto et~al.}(1985)\citenamefont{Kroto, , Heath,
  O'Brien, Curl, and Smalley}}]{kroto1985}
\bibinfo{author}{\bibfnamefont{H.~W.} \bibnamefont{Kroto}}, ,
  \bibinfo{author}{\bibfnamefont{J.~R.} \bibnamefont{Heath}},
  \bibinfo{author}{\bibfnamefont{S.~C.} \bibnamefont{O'Brien}},
  \bibinfo{author}{\bibfnamefont{R.~F.} \bibnamefont{Curl}}, \bibnamefont{and}
  \bibinfo{author}{\bibfnamefont{R.~E.} \bibnamefont{Smalley}},
  \bibinfo{journal}{Nature} \textbf{\bibinfo{volume}{318}},
  \bibinfo{pages}{162} (\bibinfo{year}{1985}).

\bibitem[{\citenamefont{Kr\"atschmer et~al.}(1990)\citenamefont{Kr\"atschmer,
  Lamb, Fostiropoulos, and Huffman}}]{kratschmer1990}
\bibinfo{author}{\bibfnamefont{W.}~\bibnamefont{Kr\"atschmer}},
  \bibinfo{author}{\bibfnamefont{L.~D.} \bibnamefont{Lamb}},
  \bibinfo{author}{\bibfnamefont{K.}~\bibnamefont{Fostiropoulos}},
  \bibnamefont{and} \bibinfo{author}{\bibfnamefont{D.~R.}
  \bibnamefont{Huffman}}, \bibinfo{journal}{Nature}
  \textbf{\bibinfo{volume}{347}}, \bibinfo{pages}{354 } (\bibinfo{year}{1990}).

\bibitem[{\citenamefont{Howard et~al.}(1991)\citenamefont{Howard, McKinnon,
  Makarovsky, Lafleur, and Johnson}}]{howard1991}
\bibinfo{author}{\bibfnamefont{J.~B.} \bibnamefont{Howard}},
  \bibinfo{author}{\bibfnamefont{J.~T.} \bibnamefont{McKinnon}},
  \bibinfo{author}{\bibfnamefont{Y.}~\bibnamefont{Makarovsky}},
  \bibinfo{author}{\bibfnamefont{A.~L.} \bibnamefont{Lafleur}},
  \bibnamefont{and} \bibinfo{author}{\bibfnamefont{M.~E.}
  \bibnamefont{Johnson}}, \bibinfo{journal}{Nature}
  \textbf{\bibinfo{volume}{352}}, \bibinfo{pages}{139} (\bibinfo{year}{1991}).

\bibitem[{\citenamefont{Homann}(1998)}]{homann1998}
\bibinfo{author}{\bibfnamefont{K.-H.} \bibnamefont{Homann}},
  \bibinfo{journal}{Angew. Chem. Int. Ed.} \textbf{\bibinfo{volume}{37}},
  \bibinfo{pages}{2434} (\bibinfo{year}{1998}).

\bibitem[{\citenamefont{Mckinnon et~al.}(1992)\citenamefont{Mckinnon, Bell, and
  Barkley}}]{mckinnon1992}
\bibinfo{author}{\bibfnamefont{J.~T.} \bibnamefont{Mckinnon}},
  \bibinfo{author}{\bibfnamefont{W.~L.} \bibnamefont{Bell}}, \bibnamefont{and}
  \bibinfo{author}{\bibfnamefont{R.~M.} \bibnamefont{Barkley}},
  \bibinfo{journal}{Combust. Flame} \textbf{\bibinfo{volume}{88}},
  \bibinfo{pages}{102} (\bibinfo{year}{1992}).

\bibitem[{\citenamefont{Osawa et~al.}(1997)\citenamefont{Osawa, Hirose, Kimura,
  Shibuya, Gu, and Li}}]{osawa1997}
\bibinfo{author}{\bibfnamefont{E.}~\bibnamefont{Osawa}},
  \bibinfo{author}{\bibfnamefont{Y.}~\bibnamefont{Hirose}},
  \bibinfo{author}{\bibfnamefont{A.}~\bibnamefont{Kimura}},
  \bibinfo{author}{\bibfnamefont{M.}~\bibnamefont{Shibuya}},
  \bibinfo{author}{\bibfnamefont{Z.-N.} \bibnamefont{Gu}}, \bibnamefont{and}
  \bibinfo{author}{\bibfnamefont{F.-M.} \bibnamefont{Li}},
  \bibinfo{journal}{Fullerenes, Nanotubes, Carbon Nanostruct.}
  \textbf{\bibinfo{volume}{5}}, \bibinfo{pages}{177} (\bibinfo{year}{1997}).

\bibitem[{\citenamefont{Becker et~al.}(1994)\citenamefont{Becker, Bada, Winans,
  Hunt, Bunch, and French}}]{becker1994}
\bibinfo{author}{\bibfnamefont{L.}~\bibnamefont{Becker}},
  \bibinfo{author}{\bibfnamefont{J.}~\bibnamefont{Bada}},
  \bibinfo{author}{\bibfnamefont{R.}~\bibnamefont{Winans}},
  \bibinfo{author}{\bibfnamefont{J.}~\bibnamefont{Hunt}},
  \bibinfo{author}{\bibfnamefont{T.}~\bibnamefont{Bunch}}, \bibnamefont{and}
  \bibinfo{author}{\bibfnamefont{B.}~\bibnamefont{French}},
  \bibinfo{journal}{Science} \textbf{\bibinfo{volume}{265}},
  \bibinfo{pages}{642} (\bibinfo{year}{1994}).

\bibitem[{\citenamefont{Sadana et~al.}(2005)\citenamefont{Sadana, Liang,
  Brinson, Arepalli, Farhat, Hauge, Smalley, and Billups}}]{sadana2005}
\bibinfo{author}{\bibfnamefont{A.}~\bibnamefont{Sadana}},
  \bibinfo{author}{\bibfnamefont{F.}~\bibnamefont{Liang}},
  \bibinfo{author}{\bibfnamefont{B.}~\bibnamefont{Brinson}},
  \bibinfo{author}{\bibfnamefont{S.}~\bibnamefont{Arepalli}},
  \bibinfo{author}{\bibfnamefont{S.}~\bibnamefont{Farhat}},
  \bibinfo{author}{\bibfnamefont{R.}~\bibnamefont{Hauge}},
  \bibinfo{author}{\bibfnamefont{R.}~\bibnamefont{Smalley}}, \bibnamefont{and}
  \bibinfo{author}{\bibfnamefont{W.}~\bibnamefont{Billups}},
  \bibinfo{journal}{J. Phys. Chem. B} \textbf{\bibinfo{volume}{109}},
  \bibinfo{pages}{4416} (\bibinfo{year}{2005}), ISSN \bibinfo{issn}{1520-6106}.

\bibitem[{\citenamefont{Smalley}(1992)}]{smalley1992}
\bibinfo{author}{\bibfnamefont{R.~E.} \bibnamefont{Smalley}},
  \bibinfo{journal}{Acc. Chem. Res} \textbf{\bibinfo{volume}{25}},
  \bibinfo{pages}{98} (\bibinfo{year}{1992}).

\bibitem[{\citenamefont{Heath}(1998)}]{heath1998}
\bibinfo{author}{\bibfnamefont{J.~R.} \bibnamefont{Heath}},
  \bibinfo{journal}{Nature} \textbf{\bibinfo{volume}{393}},
  \bibinfo{pages}{730} (\bibinfo{year}{1998}).

\bibitem[{\citenamefont{v.~Helden et~al.}(1993)\citenamefont{v.~Helden, Gotts,
  and Bowers}}]{helden1993}
\bibinfo{author}{\bibfnamefont{G.}~\bibnamefont{v.~Helden}},
  \bibinfo{author}{\bibfnamefont{N.~G.} \bibnamefont{Gotts}}, \bibnamefont{and}
  \bibinfo{author}{\bibfnamefont{M.~T.} \bibnamefont{Bowers}},
  \bibinfo{journal}{Nature} \textbf{\bibinfo{volume}{363}}, \bibinfo{pages}{60}
  (\bibinfo{year}{1993}).

\bibitem[{\citenamefont{Hunter et~al.}(1993)\citenamefont{Hunter, Fye, and
  Jarrold}}]{hunter1993}
\bibinfo{author}{\bibfnamefont{J.}~\bibnamefont{Hunter}},
  \bibinfo{author}{\bibfnamefont{J.}~\bibnamefont{Fye}}, \bibnamefont{and}
  \bibinfo{author}{\bibfnamefont{M.~F.} \bibnamefont{Jarrold}},
  \bibinfo{journal}{Science} \textbf{\bibinfo{volume}{260}},
  \bibinfo{pages}{784} (\bibinfo{year}{1993}).

\bibitem[{\citenamefont{Irle et~al.}(2006)\citenamefont{Irle, Zheng, Wang, and
  Morokuma}}]{irle2006}
\bibinfo{author}{\bibfnamefont{S.}~\bibnamefont{Irle}},
  \bibinfo{author}{\bibfnamefont{G.}~\bibnamefont{Zheng}},
  \bibinfo{author}{\bibfnamefont{Z.}~\bibnamefont{Wang}}, \bibnamefont{and}
  \bibinfo{author}{\bibfnamefont{K.}~\bibnamefont{Morokuma}},
  \bibinfo{journal}{J. Phys. Chem. B} \textbf{\bibinfo{volume}{110}},
  \bibinfo{pages}{14531} (\bibinfo{year}{2006}), ISSN
  \bibinfo{issn}{1520-6106}.

\bibitem[{\citenamefont{Huang et~al.}(2007)\citenamefont{Huang, Ding, Jiao, and
  Yakobson}}]{huang2007}
\bibinfo{author}{\bibfnamefont{J.~Y.} \bibnamefont{Huang}},
  \bibinfo{author}{\bibfnamefont{F.}~\bibnamefont{Ding}},
  \bibinfo{author}{\bibfnamefont{K.}~\bibnamefont{Jiao}}, \bibnamefont{and}
  \bibinfo{author}{\bibfnamefont{B.~I.} \bibnamefont{Yakobson}},
  \bibinfo{journal}{Phys. Rev. Lett.} \textbf{\bibinfo{volume}{99}},
  \bibinfo{eid}{175503} (pages~\bibinfo{numpages}{4}) (\bibinfo{year}{2007}).

\bibitem[{\citenamefont{Singh and Srivastava}(1995)}]{singh1995}
\bibinfo{author}{\bibfnamefont{H.}~\bibnamefont{Singh}} \bibnamefont{and}
  \bibinfo{author}{\bibfnamefont{M.}~\bibnamefont{Srivastava}},
  \bibinfo{journal}{Energy Sources A} \textbf{\bibinfo{volume}{17}},
  \bibinfo{pages}{615} (\bibinfo{year}{1995}).

\bibitem[{\citenamefont{Goroff}(29)}]{goroff1996}
\bibinfo{author}{\bibfnamefont{N.}~\bibnamefont{Goroff}},
  \bibinfo{journal}{Acc. Chem. Res.} \textbf{\bibinfo{volume}{1996}},
  \bibinfo{pages}{77} (\bibinfo{year}{29}).

\bibitem[{\citenamefont{Lozovik and Popov}(1997)}]{lozovik1997}
\bibinfo{author}{\bibfnamefont{Y.~E.} \bibnamefont{Lozovik}} \bibnamefont{and}
  \bibinfo{author}{\bibfnamefont{A.~M.} \bibnamefont{Popov}},
  \bibinfo{journal}{Physics-Uspekhi} \textbf{\bibinfo{volume}{7}},
  \bibinfo{pages}{751} (\bibinfo{year}{1997}).

\bibitem[{\citenamefont{Morokuma}(2007)}]{morokuma2007}
\bibinfo{author}{\bibfnamefont{K.}~\bibnamefont{Morokuma}},
  \bibinfo{journal}{Bull. Chem. Soc. Jpn.} \textbf{\bibinfo{volume}{80}},
  \bibinfo{pages}{2247} (\bibinfo{year}{2007}).

\bibitem[{\citenamefont{Feller et~al.}(1997)\citenamefont{Feller, Yin, Pastor,
  and MacKerell}}]{mackerell1997}
\bibinfo{author}{\bibfnamefont{S.~E.} \bibnamefont{Feller}},
  \bibinfo{author}{\bibfnamefont{D.}~\bibnamefont{Yin}},
  \bibinfo{author}{\bibfnamefont{R.~W.} \bibnamefont{Pastor}},
  \bibnamefont{and} \bibinfo{author}{\bibfnamefont{A.~D.~J.}
  \bibnamefont{MacKerell}}, \bibinfo{journal}{Biophys. J.}
  \textbf{\bibinfo{volume}{73}}, \bibinfo{pages}{2269} (\bibinfo{year}{1997}).

\bibitem[{\citenamefont{Fowler and Manolopoulos}(1995)}]{c60_bondlengths}
\bibinfo{author}{\bibfnamefont{P.~W.} \bibnamefont{Fowler}} \bibnamefont{and}
  \bibinfo{author}{\bibfnamefont{D.~E.} \bibnamefont{Manolopoulos}},
  \emph{\bibinfo{title}{An Atlas of Fullerenes}} (\bibinfo{publisher}{Oxford
  Univ. Press}, \bibinfo{year}{1995}).

\bibitem[{\citenamefont{Jursic}(1999)}]{bond_dissoc_energy}
\bibinfo{author}{\bibfnamefont{B.~S.} \bibnamefont{Jursic}},
  \bibinfo{journal}{J. Chem. Soc. Perkin Trans.} \textbf{\bibinfo{volume}{2}},
  \bibinfo{pages}{369} (\bibinfo{year}{1999}).

\bibitem[{\citenamefont{Leimkuhler et~al.}(1996)\citenamefont{Leimkuhler,
  Reich, and Skeel}}]{leapfrog}
\bibinfo{author}{\bibfnamefont{B.}~\bibnamefont{Leimkuhler}},
  \bibinfo{author}{\bibfnamefont{S.}~\bibnamefont{Reich}}, \bibnamefont{and}
  \bibinfo{author}{\bibfnamefont{R.~D.} \bibnamefont{Skeel}},
  \emph{\bibinfo{title}{IMA Volumes in Mathematics and its Applications, Vol.
  82}} (\bibinfo{publisher}{Springer-Verlag}, \bibinfo{year}{1996}).

\bibitem[{\citenamefont{van Kampen}(1981)}]{kampen1981}
\bibinfo{author}{\bibfnamefont{N.~G.} \bibnamefont{van Kampen}},
  \emph{\bibinfo{title}{Stochastic Processes in Physics and Chemistry}}
  (\bibinfo{publisher}{North-Holland}, \bibinfo{year}{1981}).

\bibitem[{\citenamefont{Bussi and Parrinello}(2007)}]{bussi2007}
\bibinfo{author}{\bibfnamefont{G.}~\bibnamefont{Bussi}} \bibnamefont{and}
  \bibinfo{author}{\bibfnamefont{M.}~\bibnamefont{Parrinello}},
  \bibinfo{journal}{Phys. Rev. E} \textbf{\bibinfo{volume}{75}},
  \bibinfo{eid}{056707} (pages~\bibinfo{numpages}{7}) (\bibinfo{year}{2007}).

\bibitem[{\citenamefont{Binggeli and Chelikowsky}(1994)}]{chelikowsky1996}
\bibinfo{author}{\bibfnamefont{N.}~\bibnamefont{Binggeli}} \bibnamefont{and}
  \bibinfo{author}{\bibfnamefont{J.~R.} \bibnamefont{Chelikowsky}},
  \bibinfo{journal}{Phys. Rev. B} \textbf{\bibinfo{volume}{50}},
  \bibinfo{pages}{11764} (\bibinfo{year}{1994}).

\bibitem[{\citenamefont{Xu and Scuseria}(1994)}]{scuseria1994}
\bibinfo{author}{\bibfnamefont{C.}~\bibnamefont{Xu}} \bibnamefont{and}
  \bibinfo{author}{\bibfnamefont{G.~E.} \bibnamefont{Scuseria}},
  \bibinfo{journal}{Phys. Rev. Lett.} \textbf{\bibinfo{volume}{72}},
  \bibinfo{pages}{669} (\bibinfo{year}{1994}).

\bibitem[{\citenamefont{Leach}(2001)}]{leach2001}
\bibinfo{author}{\bibfnamefont{A.}~\bibnamefont{Leach}},
  \emph{\bibinfo{title}{Molecular Modelling: Principles and Applications}}
  (\bibinfo{publisher}{Pearson Education}, \bibinfo{year}{2001}),
  \bibinfo{note}{iSBN:0582382106}.

\bibitem[{\citenamefont{Huczko et~al.}(1997)\citenamefont{Huczko, Lange,
  Byszewski, Poplawska, and Starski}}]{huczko1997}
\bibinfo{author}{\bibfnamefont{A.}~\bibnamefont{Huczko}},
  \bibinfo{author}{\bibfnamefont{H.}~\bibnamefont{Lange}},
  \bibinfo{author}{\bibfnamefont{P.}~\bibnamefont{Byszewski}},
  \bibinfo{author}{\bibfnamefont{M.}~\bibnamefont{Poplawska}},
  \bibnamefont{and} \bibinfo{author}{\bibfnamefont{A.}~\bibnamefont{Starski}},
  \bibinfo{journal}{J. Phys. Chem.} \textbf{\bibinfo{volume}{101}},
  \bibinfo{pages}{1267} (\bibinfo{year}{1997}).

\bibitem[{\citenamefont{S\"aidane et~al.}(2004)\citenamefont{S\"aidane,
  Razafinimanana, Lange, Huczko, Baltas, Gleizes, and Meunier}}]{saidane2004}
\bibinfo{author}{\bibfnamefont{K.}~\bibnamefont{S\"aidane}},
  \bibinfo{author}{\bibfnamefont{M.}~\bibnamefont{Razafinimanana}},
  \bibinfo{author}{\bibfnamefont{H.}~\bibnamefont{Lange}},
  \bibinfo{author}{\bibfnamefont{A.}~\bibnamefont{Huczko}},
  \bibinfo{author}{\bibfnamefont{M.}~\bibnamefont{Baltas}},
  \bibinfo{author}{\bibfnamefont{A.}~\bibnamefont{Gleizes}}, \bibnamefont{and}
  \bibinfo{author}{\bibfnamefont{J.-L.} \bibnamefont{Meunier}},
  \bibinfo{journal}{J. Phys. D: Appl. Phys.} \textbf{\bibinfo{volume}{37}},
  \bibinfo{pages}{232} (\bibinfo{year}{2004}).

\bibitem[{\citenamefont{Akita et~al.}(2000)\citenamefont{Akita, Ashihara, and
  Nakayama}}]{akita2000}
\bibinfo{author}{\bibfnamefont{S.}~\bibnamefont{Akita}},
  \bibinfo{author}{\bibfnamefont{H.}~\bibnamefont{Ashihara}}, \bibnamefont{and}
  \bibinfo{author}{\bibfnamefont{Y.}~\bibnamefont{Nakayama}},
  \bibinfo{journal}{Jpn. J. Appl. Phys.} \textbf{\bibinfo{volume}{39}},
  \bibinfo{pages}{4939} (\bibinfo{year}{2000}).

\bibitem[{\citenamefont{Markovic et~al.}(2003)\citenamefont{Markovic,
  Todorovic-Markovic, Marinkovic, and Nenadovic}}]{markovic2003}
\bibinfo{author}{\bibfnamefont{Z.}~\bibnamefont{Markovic}},
  \bibinfo{author}{\bibfnamefont{B.}~\bibnamefont{Todorovic-Markovic}},
  \bibinfo{author}{\bibfnamefont{M.}~\bibnamefont{Marinkovic}},
  \bibnamefont{and}
  \bibinfo{author}{\bibfnamefont{T.}~\bibnamefont{Nenadovic}},
  \bibinfo{journal}{Carbon} \textbf{\bibinfo{volume}{41}}, \bibinfo{pages}{369}
  (\bibinfo{year}{2003}).

\bibitem[{\citenamefont{Markovic et~al.}(2002)\citenamefont{Markovic,
  Todorovic-Markovic, and Nenadovic}}]{markovic2002}
\bibinfo{author}{\bibfnamefont{Z.}~\bibnamefont{Markovic}},
  \bibinfo{author}{\bibfnamefont{B.}~\bibnamefont{Todorovic-Markovic}},
  \bibnamefont{and}
  \bibinfo{author}{\bibfnamefont{T.}~\bibnamefont{Nenadovic}},
  \bibinfo{journal}{Fullerenes, Nanotubes, Carbon Nanostruct.}
  \textbf{\bibinfo{volume}{10}}, \bibinfo{pages}{81} (\bibinfo{year}{2002}).

\end{thebibliography}
\end{document}